\documentclass[prc,twocolumn,showpacs,preprintnumbers,superscriptaddress,
               amsmath,amssymb]{revtex4}
\usepackage{amsmath}           
\usepackage{amsfonts,mathrsfs}
\usepackage{graphicx}          
\usepackage{dcolumn}           

\begin{document}

\preprint{\fbox{\sc version of \today}}

\title{Nuclear Halos and Drip Lines in Symmetry-Conserving Continuum HFB Theory.}
\author{N. Schunck}
\affiliation{Departamento de Fisica Teorica, Universidad Autonoma de Madrid, 28 049 Cantoblanco, Madrid, Spain}
\affiliation{Department of Physics and Astronomy, University of Tennessee, Knoxville, Tennessee 37996, USA}
\affiliation{Physics Division, Oak Ridge National Laboratory, P.O. Box 2008, Oak Ridge, Tennessee 37831, USA}

\author{J. L. Egido}
\affiliation{Departamento de Fisica Teorica, Universidad Autonoma de Madrid, 28 049 Cantoblanco, Madrid, Spain}

\date{\today}

\begin{abstract}
We review the properties of nuclear halos and nuclear skins in drip line nuclei in the framework of the spherical Hartree-Fock-Bogoliubov theory with continuum effects and projection on good particle number with the Gogny force. We first establish the position of the un-projected HFB drip lines for the two most employed parametrizations of the Gogny force and show that the use of finite-range interactions leads almost always to small-sized halos, even in the least bound nuclei, which is in agreement with most mean-field predictions. We also discuss the size of the neutron skin at the drip line and its relation to neutron asymmetry. The impact of particle-number projection and its conceptual consequences near the drip line are analyzed in detail. In particular, we discuss the role of the chemical potential in a projected theory and the criteria required to define the drip line. We show that including particle number projection can shift the latter, in particular near closed shells. We notice that, as a result, the size of the halo can be increased due to larger pairing correlations. However, combining the most realistic pairing interaction, a proper treatment of the continuum and particle number projection does not permit to reproduce the very large halos observed in very light nuclei.
\end{abstract}  


\maketitle

Neutron-rich nuclei present us with unique opportunities to test nuclear models. As the asymmetry between the number of neutrons and protons in atomic nuclei increases, a number of new phenomena appear such as neutron skins, nuclear halos or shell-melting \cite{DripLineReview-1,DripLineReview-2}. With current and on-going developments of radioactive ion beam facilities, new territories of the nuclear chart are open to exploration and data from very neutron-rich nuclei is likely to upset the existing theories \cite{MgNature}.

The so-called nuclear skin in neutron-rich nuclei is caused to a large extent by the isospin asymmetry. In the mean-field picture of the atomic nucleus, it should therefore depend mostly on the iso-vector component of the effective hamiltonian or Lagrangian. By contrast, the understanding of nuclear halos in the context of mean-field theories is quite fragmentary and it is not clear if one can relate this phenomenon to a particular term of the effective interaction or density functional. The extreme difficulties to approach the drip lines in heavy nuclei, where mean-field theories are most often employed, also forbids to test the calculations against experiment. In light nuclei, where experimental data is available, few-body models that introduce a core surrounded by one (two-body models) or two particles (three-body models) are very successful \cite{Halo-review1,Halo-review2}. However mean-field models are notoriously unreliable in these extremely light systems unless severe corrections beyond the mean-field are included. 

In heavy nuclei it is commonly thought that nuclear halos should be interpreted as resulting from the coupling to the continuum via residual interactions such as pairing correlations \cite{BertschHalo}. The spatial delocalization of continuum states gives a simple motivation for such interpretations. Nevertheless mean-field theories show significant variations in their predictions of halos. Several major difficulties can explain these discrepancies: firstly, the neutron drip line is not known beyond the Oxygen element. This lack of experimental data is going to be partially filled in the near future but imply that models can not really be benchmarked against experiment. Secondly all self-consistent approaches to nuclear structure rely on the parametrization of some effective interaction, Lagrangian, or energy density functional. The extrapolability of such interactions to regions of very large neutron excess is by no means guaranteed by the theory. Furthermore, beyond mean-field effects such as symmetry restoration mechanisms or configuration mixing might be playing a different role in these extreme regions as in the valley of stability. 

In a recent article, we proposed  \cite{nous} a simple and effective method to treat the continuum in configuration space for mean-field theories based on finite-range interactions of the Gogny type. We showed that our procedure provides the correct asymptotic behavior of the nuclear wave-functions and established the proton and neutron drip lines with the D1S interaction. Restoration of the particle number symmetry was also discussed and it was found that in some cases, in particular when the neutron number is near a closed shell, the variation after projection (VAP) method could shift the position of the drip lines by 2 neutrons.
In this paper, we apply our method to the specific problems of nuclear skins and nuclear halos near the neutron drip line. We will furthermore review several cases that have been proposed in the past, either in the framework of the Skyrme Hartree-Fock Bogoliubov theory or the relativistic mean-field  and discuss the impact both of finite-range interaction and symmetry restoration effects. To our knowledge this constitutes the first attempt to address this problem in explicitly symmetry-conserving mean-field approach. In section I we briefly recall the main features of our approach as well as the Helm model that has been traditionally used to describe nuclear halos. In section II we discuss the nuclear skins and halos in the standard framework of the HFB theory with the Gogny interaction. In section III we investigate the impact of particle-number symmetry restoration on halos.

\section{BRIEF DESCRIPTION OF THE METHOD}

In the nuclear mean-field approach, the energy of the nucleus is computed as the expectation value of a two-body Hamiltonian on a trial wave-function \cite{RingSchuck}. Besides the relativistic approaches\cite{Ring_Rev}, there exists two main families of two-body effective forces to this date, the zero-range Skyrme \cite{skyrme_review} interaction and the finite-range Gogny one \cite{D1}. Both are empirical effective forces and there exists a number of realistic parametrizations. Skyrme forces lead to a local energy density which is the basic building block of the nuclear Energy Density Functional (EDF) theory \cite{hfb_review}. The Gogny interaction, because of its finite range, is non-local and computationally more involved, for this reason the calculations are most conveniently carried out in configuration space, i.e. the solutions to the Hartree-Fock (HF) or Hartree-Fock-Bogoliubov (HFB) problem are expanded on a given basis.

The harmonic oscillator (HO) basis has always played a special role in configuration space calculations, as its eigenfunctions are given analytically and are separable. However, a well-known deficiency of this basis is that it is made exclusively of bound-states since the underlying potential has infinite walls. In practice, since calculations are always performed in a given truncation scheme (for a fixed cut-off of the basis) the localization of all HO basis states imply that the physical wave-functions of the system will always acquire a Gaussian asymptotic, including the weakly-bound and positive-energy states. This is clearly unrealistic, as spherical continuum states should be spherical waves. The consequences of this deficiency become more serious close to the drip line, as pairing correlations can couple discrete bound-states to the continuum. It is thus critical to properly describe the continuous spectrum even at the level of ground-state calculations \cite{Doba-continuum}. 

There exists a number of techniques to take into account the continuum in nuclear structure calculations, and it is not the purpose of this article to discuss in detail the merits of every one of them. Let us just mention briefly for completeness: coordinate-space Skyrme HFB theory with either vanishing \cite{Doba-continuum} or outgoing-wave boundary conditions \cite{grasso}, coordinate-space Relativistic Hartree-Bogoliubov with finite-element method \cite{RHB_FiniteElem}, the use of the Gamow basis in the Skyrme HFB theory  \cite{GamowHFB}, in the Continuum Shell Model \cite{SMEC} and in the Gamow Shell Model \cite{GamowSM}. At the present time technical difficulties in including the full continuum with the exact resonant and non-resonant spectrum lead to the consequence that the most advanced theories are only applied with simple model interactions that are tailored to capture the main physical properties of the system. Only in the coordinate-space HFB approach realistic Skyrme interactions were employed with density-dependence zero-range forces in the pairing channel (requiring the introduction of a cut-off in the quasi-particle spectrum or a regularization procedure \cite{regularization}). Moreover, as far as mean-field based theories are concerned, no attempt has been done to include with the coupling to the continuum the restoration of broken symmetries or collective motion.

Therefore, in order to combine the flexibility of configuration space calculations with the necessary inclusion of the continuum, it has been proposed in \cite{ZMR.03,nous} to work in a basis made of the eigenstates of the Woods-Saxon potential. The latter are obtained by integrating the Schr\"{o}dinger equation in a box of size $R_{box}$ with a mesh size $h$. In practice $R_{box} = 20$ fm and $h = 0.1$ fm are sufficient to obtain a good convergence of the solutions. Boundary conditions are set on the walls of the box. As usual, several choices are possible. Outgoing wave boundary conditions lead to wave-functions that are not square-integrable, and special techniques must be employed to overcome this difficulty \cite{nonHermitian-1,nonHermitian-2,nonHermitian-3,nonHermitian-4}. Vanishing boundary conditions guarantee that the basis functions are square-integrable and can thus be normalized at the price of eliminating all the continuum states that do not have a node on the walls of the box. It was shown in \cite{BoxVsGamow} that both techniques essentially lead to very similar results as far as bound-states and bulk properties of nuclei are concerned. In the following, we use vanishing box boundary conditions.

In our calculations we use  the finite-range Gogny interaction \cite{D1}. The same interaction is used in the particle-hole channel (mean-field) and particle-particle channel (pairing) and both the direct and exchange contributions coming from {\it all} the terms of the interaction are taken into account in the calculation. The finite-range of the force in the pairing channel allows to avoid the divergence problem (in momentum space) and cut-off dependence of zero-range forces. All of our calculations are performed in spherical symmetry.

To obtain quantitative information on the neutron halo in neutron-rich nuclei we make use of the Helm method \cite{Helm1,Helm2,Helm3,HelmDoba}. Firstly, the neutrons (protons) form factor is computed as the Fourier transform of the neutrons (protons) density. In spherical symmetry, this leads to:
\begin{equation}
F(q) = 4\pi\int_{0}^{\infty} j_{0}(qr)\rho_{\tau}(r)r^{2}dr
\label{formfactor}
\end{equation}
where $q$ is the momentum, $j_{0}$ is the spherical Bessel function of order 0 and $\rho_{\tau}(r)$ is the density ($\tau$ standing for neutron or proton). This form factor built out of the realistic one-body density, in our case calculated with the 
Gogny force, is then compared to the Helm form-factor obtained from the convolution of the Gaussian profile 
\begin{equation}
f_G(r) = \frac{e^{{-r^2/(2\sigma^2)}}}{(2\pi)^{3/2}\sigma^3}
\end{equation}
with a sharp density profile: $\rho(r) = \rho_{0}$ for $r \leq R_{0}$ and $\rho(r) = 0$ elsewhere. Since this model is presented in details in the references quoted, we simply recall the formulas we are going to use.  The two parameters $R_0$ and $\sigma$ of the model are determined in the following way. 

The rms radius $R_{rms}$ is defined as the squared root of the mean-value of the operator $\hat{r}^{2}$. It is extracted from the nucleonic density:
\begin{equation}
R_{rms} = \sqrt{\langle \hat{r}^{2}\rangle} = 
\sqrt{\frac{\displaystyle\int d^{3}\vec{r}\;r^{2}\rho(\vec{r})}{\displaystyle\int d^{3}\vec{r}\;\rho(\vec{r})}}
\end{equation}
For the Helm radius one straightforwardly obtains 
\begin{equation}
R_{rms}^{H} = \sqrt{\frac{3}{5}(R_{0}^{2} + 5\sigma^{2})}
\end{equation}
where $R_{0}$ is the diffraction radius:
\begin{equation}
R_{0} = 4.49341/q_{1}
\end{equation}
and $q_{1}$ is the first zero of the realistic form-factor (\ref{formfactor}) obtained in our theoretical approach. The surface thickness $\sigma$ is defined as:
\begin{equation}
\sigma^{2} = \frac{2}{q_{m}^{2}}\ln \frac{3N j_{1}(q_{m}R_{0})}{R_{0}q_{m}F(q_{m})}
\end{equation}
where $N$ is the number of particles, $j_{1}$ the spherical Bessel function of order 1 and $q_{m}$ is the first maximum of the realistic form-factor (\ref{formfactor}). 

At this point we should note that the method  does not carry out any information on the eventual decorrelation between a core and a few valence particles. It only provides a simple and fast method to assess the spatial extension of the nucleus and an excellent  starting point to determine the best halo candidates. However, in few-body nuclear models, the nuclear halo is often interpreted as one single nucleon or a pair of nucleons orbiting around a core, see e.g. \cite{Halo-2body-1,Halo-2body-2} for two-body models and \cite{Halo-3body-1,Halo-3body-2,Halo-3body-3} for 3-body models. In order to reconcile these cluster approaches with a mean-field description of the nucleus, a more detailed analysis of the density should be carried out. Alternative techniques have been proposed to cure this deficiency \cite{HaloDuguet}.

It is usually convenient to multiply the rms and the Helm radius by $\sqrt{5/3}$. The quantities
\begin{equation}
R_{geom} = R_{rms}\sqrt{\frac{5}{3}}; \hspace{0.5cm}R_{Helm} = R_{rms}^{H}\sqrt{\frac{5}{3}}.
\end{equation}
are related to the underlying shape of the nucleus.
A measure of the nuclear halo is then provided by the quantity:
\begin{equation}
\delta R_{halo} = R_{geom} - R_{Helm}
\label{deltaR}
\end{equation}
The neutron skin can be defined in various ways depending on which type of radius is considered. Its general expression is:
\begin{equation}
\Delta R = R^{(n)} - R^{(p)}
\end{equation}
where $R$ can be either the geometrical radius, the Helm radius or the diffraction radius. It was argued in \cite{HelmDoba} that the best approximation to the neutron skin is obtained when taking the Helm radius, as the latter is somewhat rid of spurious contributions coming from the neutron halo.

\section{NUCLEAR SKINS AND HALOS WITH FINITE-RANGE INTERACTIONS}

Systematic calculations near the neutron drip line have been carried out using the spherical HFB code in the Woods-Saxon basis that was presented in \cite{nous}. The basis was constructed from the eigenstates of the WS potential with the universal parametrization of \cite{WSUniversal} applied to the $Z=126$ and $N=184$ nucleus. The Schr\"{o}dinger equation was integrated in a box of $R_{box} = 20$ fm with vanishing boundary conditions. All eigenstates with $\ell \leq 15$ and $n\leq 18$ were retained in the basis. As was shown in \cite{nous}, such a choice guarantees a good convergence of the subsequent HFB calculation.

\subsection{Determination and properties of the neutron drip line}

There exist few parametrizations of the Gogny interaction: in our calculations we considered the parametrizations D1 of \cite{D1} and D1S of \cite{D1S}. For each of them the neutron drip line was calculated based on the requirement that the one-neutron $S_{n} = B(N,Z) - B(N-1,Z)$ separation energy must be negative for bound nuclei. Since $S_{2n} = S_{n} + S_{n-1}$, the criterion $S_{n} < 0$ is stricter than the condition that the two-neutron separation energy is negative. In the HFB theory the one-neutron separation energy $S_n$ is approximated by the neutrons Fermi energy $\lambda_n = dE/dN \approx -S_{n} $. A nearly equivalent condition to define the one neutron drip line is therefore $\lambda_n > 0$. When HFB pairing correlations vanish (case of closed shells), the value of the chemical potential $\lambda$ is meaningless and can not be used to define the drip line any more (Hartree-Fock limit). However, in the HF approach and within the approximation of the validity of Koopmans' theorem \cite{Koopman}, the stability of a nucleus is simply governed by the position of the last occupied level: if it has positive energy, then the nucleus is unbound with respect to particle emission. 

We display in Table \ref{table01} the one neutron drip line nuclei obtained with both interactions. In the presence of pairing correlations the criterion  $\lambda_n > 0$ has been used. For the neutron shell closures  $N = 82$ (D1S: elements $Z = 36$ to $Z=40$, D1: elements $Z = 36$ and $Z=38$), $N = 126$ (D1S: elements $Z = 52$ to $Z=64$, D1: elements $Z=54$ to $Z=62$) and $N = 184$ (D1S: elements $Z = 80$ to $Z=92$, D1: elements $Z=80$ to $Z=90$) the neutron pairing correlations vanishes and we have to rely on Koopmans' theorem. The columns corresponding to the D1S interaction were already presented in \cite{nous} and are recalled for comparison. We would like to comment at this point that, due to some technical problems with our previous codes at the aforementioned shell closures, in \cite{nous} the drip line was predicted with two neutron less for the elements Kr, Te, Xe, Ba, Hg and Pb. We also show in Table \ref{table01} the difference in the number of neutrons between the drip line nuclei with the D1 and D1S interaction: $\Delta N = N_{drip line}(D1) - N_{drip line}(D1S)$. In general the D1 parametrization predicts a drip line with more neutrons, probably due to the fact that it provides more pairing correlations than the D1S one. Let us emphasize that all calculations performed in this work are restricted to spherical symmetry. Several of the nuclei listed in Table \ref{table01} may be deformed in their ground-state \cite{Gogny-MassTable}. Symmetry-unrestricted HFB calculations of neutron-rich nuclei would most likely shift the position of the drip line in several places.

\begin{table}[h]
\begin{center}
\caption{Table of spherical HFB one neutron drip line nuclei obtained with the D1S and D1 interactions. The columns marked $\Delta N$ represent the shift of the drip line (in number of neutrons) when using the D1 interaction compared to the D1S.}
\begin{ruledtabular}
\begin{tabular}{c|cc|cc||c|cc|cc}
Z & N & D1S &  D1 & $\Delta N$ & Z& N& D1S & D1 & $\Delta N$ \\
\hline
 6   &14  & $^{~20}$C	&  $^{~20}$C &  0 &  50& 120 &  $^{170}$Sn &  $^{168}$Sn & -2 \\
 8   &18  & $^{~26}$O	&  $^{~26}$O &  0 &  52& 126 &  $^{178}$Te &  $^{178}$Te &  0 \\
 10  &20  & $^{~30}$Ne &  $^{~30}$Ne &  0 &  54& 126 &  $^{180}$Xe &  $^{180}$Xe &  0 \\
 12  &28  & $^{~40}$Mg &  $^{~42}$Mg & +2 &  56& 126 &  $^{182}$Ba &  $^{182}$Ba &  0 \\
 14  &32  & $^{~46}$Si &  $^{~46}$Si &  0 &  58& 126 &  $^{184}$Ce &  $^{184}$Ce &  0 \\
 16  &34  & $^{~50}$S  &  $^{~52}$S  & +2 &  60& 126 &  $^{186}$Nd &  $^{186}$Nd &  0 \\
 18  &38  & $^{~56}$Ar &  $^{~58}$Ar & +2 &  62& 126 &  $^{188}$Sm &  $^{188}$Sm &  0 \\
 20  &44  & $^{~64}$Ca &  $^{~62}$Ca & -2 &  64& 126 &  $^{190}$Gd &  $^{194}$Gd & +4 \\
 22  &50  & $^{~72}$Ti &  $^{~72}$Ti &  0 &  66& 132 &  $^{198}$Dy &  $^{204}$Dy & +6 \\
 24  &52  & $^{~76}$Cr &  $^{~78}$Cr & +2 &  68& 138 &  $^{206}$Er &  $^{216}$Er & +10\\
 26  &56  & $^{~82}$Fe &  $^{~82}$Fe &  0 &  70& 150 &  $^{220}$Yb &  $^{230}$Yb & +10\\
 28  &58  & $^{~86}$Ni &  $^{~88}$Ni & +2 &  72& 168 &  $^{240}$Hf &  $^{244}$Hf & +4 \\
 30  &62  & $^{~92}$Zn &  $^{~98}$Zn & +6 &  74& 178 &  $^{252}$W  &  $^{254}$W  & +2 \\
 32  &72  & $^{104}$Ge &  $^{104}$Ge &  0 &  76& 182 &  $^{258}$Os &  $^{258}$Os &  0 \\
 34  &80  & $^{114}$Se &  $^{114}$Se &  0 &  78& 182 &  $^{260}$Pt &  $^{260}$Pt &  0 \\
 36  &82  & $^{118}$Kr &  $^{118}$Kr &  0 &  80& 184 &  $^{264}$Hg &  $^{264}$Hg &  0 \\
 38  &82  & $^{120}$Sr &  $^{120}$Sr &  0 &  82& 184 &  $^{266}$Pb &  $^{266}$Pb &  0 \\
 40  &82  & $^{122}$Zr &  $^{124}$Zr & +2 &  84& 184 &  $^{268}$Po &  $^{268}$Po &  0 \\
 42  &88  & $^{130}$Mo &  $^{130}$Mo &  0 &  86& 184 &  $^{270}$Rn &  $^{270}$Rn &  0 \\
 44  &92  & $^{136}$Ru &  $^{138}$Ru & +2 &  88& 184 &  $^{272}$Ra &  $^{272}$Ra &  0 \\
 46  &94  & $^{140}$Pd &  $^{148}$Pd & +8 &  90& 184 &  $^{274}$Th &  $^{274}$Th &  0 \\
 48  &104 & $^{152}$Cd &  $^{158}$Cd & +6 &  92& 184 &  $^{276}$U  &  $^{280}$U  & +4 \\
     &    &            &             &	  &  94& 188 &  $^{282}$Pu &  $^{294}$Pu & +12 \\
\end{tabular}
\label{table01}
\end{ruledtabular}
\vspace*{-0.5cm}
\end{center} 
\end{table}

\begin{figure}[h]
\includegraphics[height=7.0cm,width=9.0cm]{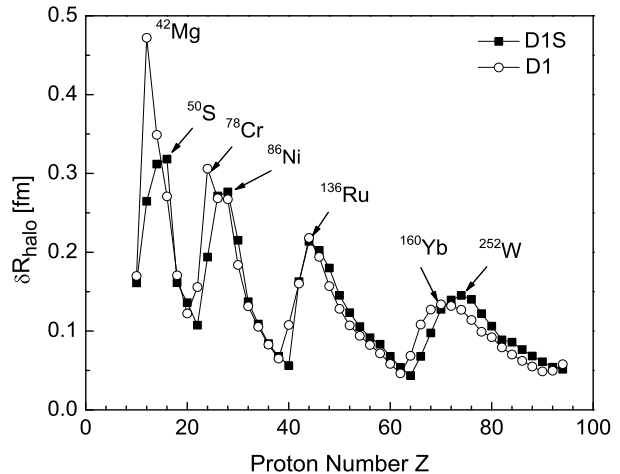}
\caption{Measure of the neutron halo: $\delta R_{halo} = R_{geom} - R_{Helm}$ in fm for spherical Gogny HFB calculations in the WS basis with the D1S (plain squares) and D1 (open circles) interactions.
	 }
\label{fig01}	
\end{figure}

For each interaction, the quantity $\delta R_{halo}$ of Eq. (\ref{deltaR}) was then computed at the drip line, i.e. for each element listed in Table \ref{table01}. The results are shown in Fig.~\ref{fig01}. We find a downward trend of $\delta R_{halo}$ as a function of $Z$ superimposed with oscillations. Both features are well understood. The decreasing behavior has to do with the well-known fact that light nuclei have larger halos. The oscillations are related to the neutron magic numbers: To the five minima (for the D1S, for example,  Z$_{\rm min}=10, 22, 40, 64 $ and 92 ) correspond to neutron number 20, 50, 82, 126 and 184, see Table \ref{table01}. For proton numbers two (four or six) units larger than a given Z$_{\rm min}$ a few neutrons occupy a new large j-shell thereby inducing pairing correlations and producing a halo. For Z values much larger than a given Z$_{\rm min}$ the number of neutrons in the shell becomes large and the halo disappears. 

As noticed in \cite{HelmDoba}, the size of the halos  is correlated with the corresponding chemical potential of the HFB solutions: Larger halos correspond to nuclei with values of $\lambda_n$ close to zero and smaller ones to large $\lambda_n$ values. As far as the effect of the parametrization of the Gogny force is concerned, we observe that the largest difference takes place in $^{42}$Mg which is not bound with D1S interaction while it is bound for the D1 interaction. Apart from this particular nucleus, both parametrizations of the Gogny force give very similar results, even though the isotopes of the drip line elements are sometimes very different, for example $^{216}$Er with the D1 interaction and $^{206}$Er with the D1S. 

It is instructive to compare our results with the work of \cite{HelmDoba}. It was pointed out in this reference that the size of the halo, as measured by the quantity $\delta R_{halo}$, significantly depends on the interaction used. A similar conclusion was reached in \cite{HaloDuguet} using a slightly different analysis procedure. In our case both parametrizations provide rather similar results in spite of the fact that the numerical values of the D1S and D1 parametrizations are quite different. It is also interesting to note that both parametrizations can lead in some cases to significantly different drip lines - in the case of the D1 interaction for example, the drip line near Palladium isotopes ($Z=46$) and Erbium ($Z=68$) and Ytterbium ($Z=70$) is located 8 and 10 neutrons further away, respectively, than in the case of the D1S interaction. Yet, as mentioned, the size of the halo remains very similar.

\begin{figure}[h]
\includegraphics[height=7.0cm,width=9.0cm]{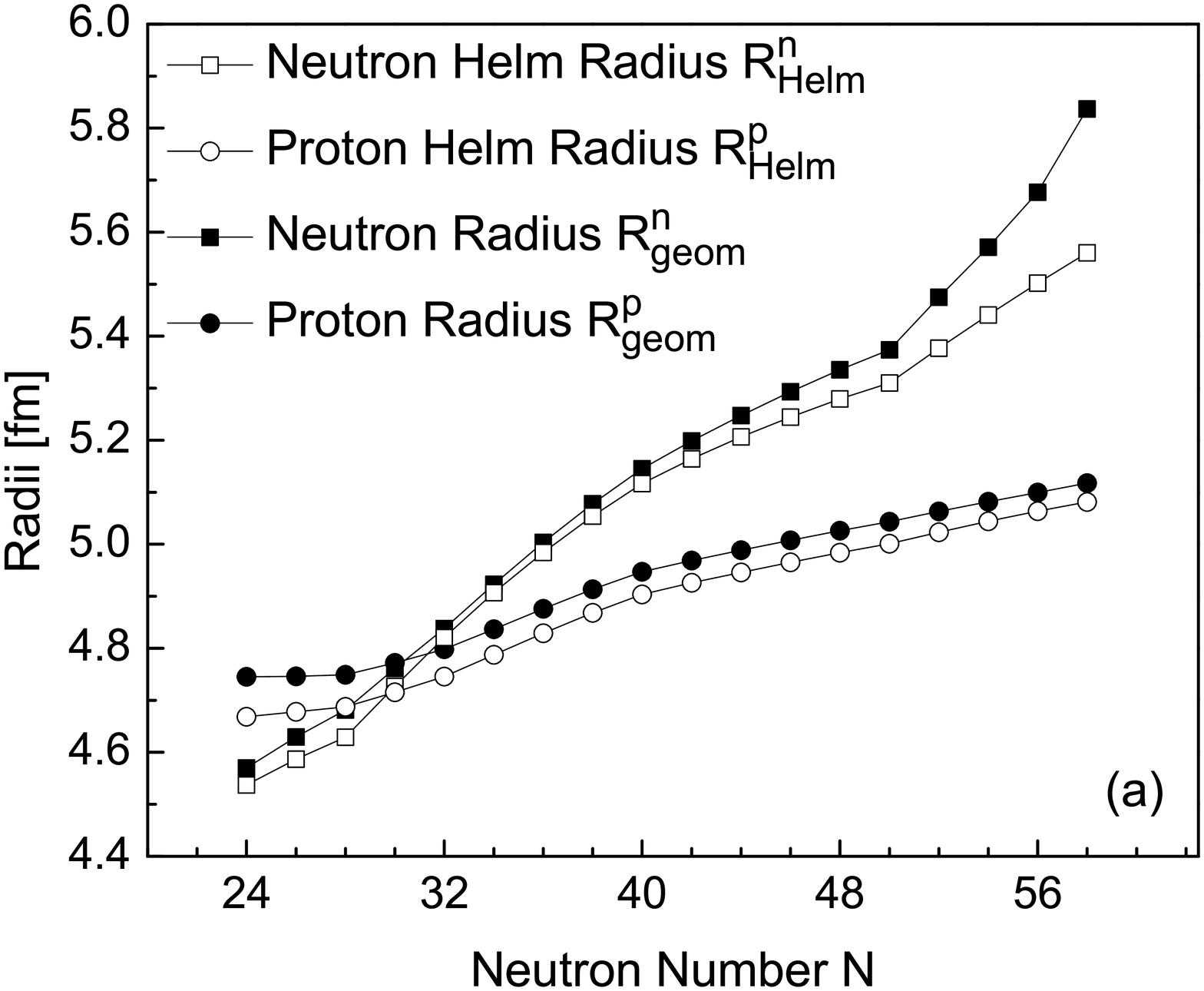}
\includegraphics[height=7.0cm,width=9.0cm]{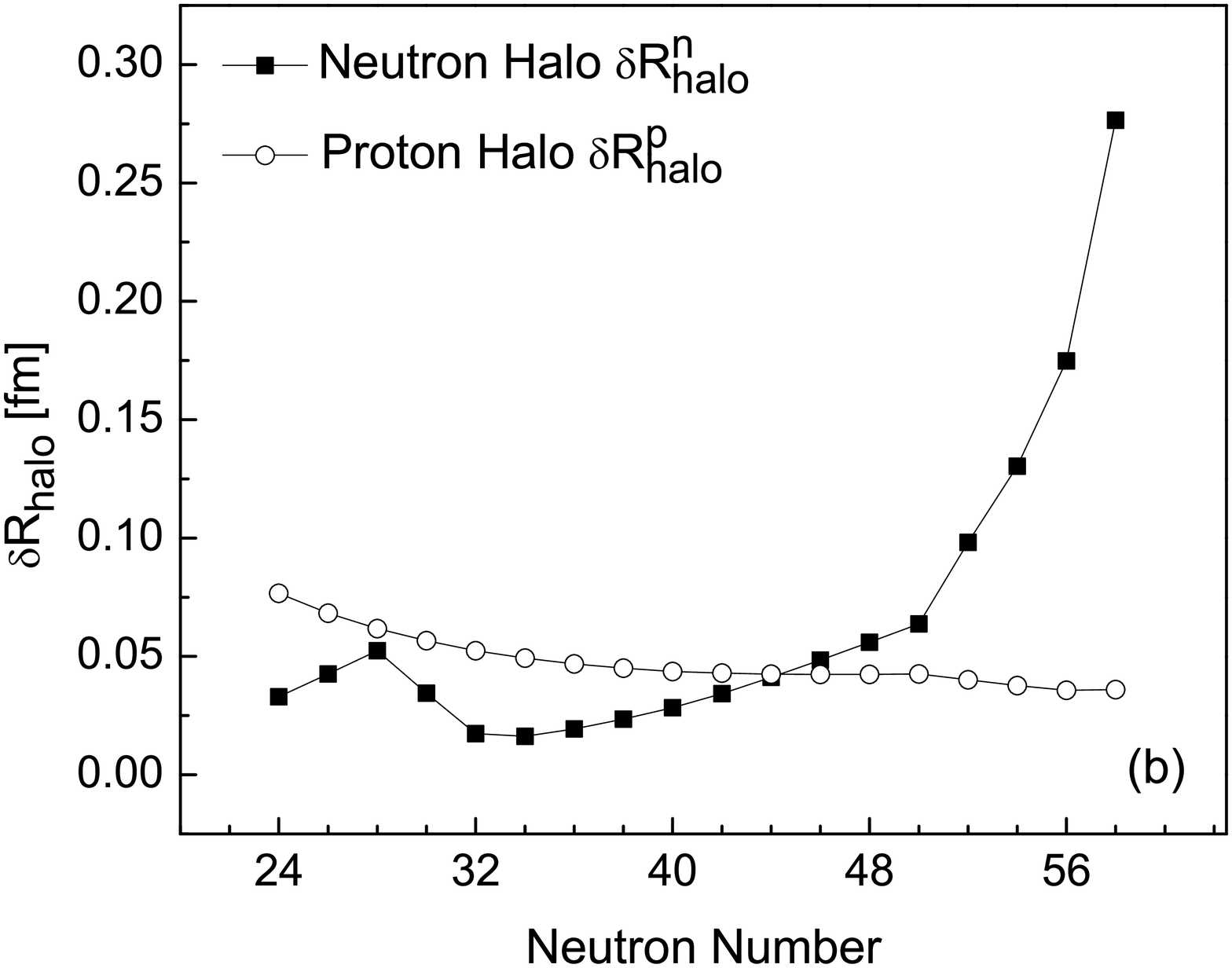}
\caption{Upper panel: Neutron $R_{geom}(n)$ and $R_{Helm}(n)$ and proton $R_{geom}(p)$ and $R_{Helm}(p)$ radii for the Ni isotopes calculated with the D1S Gogny interaction. Lower panel: Neutron and proton halo parameter $\delta R_{halo}$ along this isotopic chain.
	 }
\label{fig02}	
\end{figure}

Figure \ref{fig01} shows which elements can be considered as the best halo candidates. For each such candidate, the inspection of the full isotopic sequence from drip line to drip line can provide information on the swelling of the nuclear skin and the transition skin to halo. As a first example, we show in Fig. \ref{fig02} the case of the isotopic chain for Nickel element. For the D1S interaction, this element has one of the largest halo. Furthermore, the same isotopic line was also studied in the framework of  the Skyrme-HFB (SLy4 and SKP interactions) and RHB (NLSH and NL3 lagrangians) theories \cite{HelmDoba}, which therefore gives us results from three different sorts of mean-fields. In the upper panel of figure \ref{fig02} both the geometrical and Helm radii are plotted for the neutron and proton along the Ni isotopic chain. In the lower panel, the halo parameter $\delta R_{halo}$ is plotted for the neutrons and protons. In nuclei far away from the drip line, the difference between geometrical and Helm radius is very small, reflecting the neglegible coupling to the continuum near the valley of stability. At $N=50$ we observe the last shell closure and immediately after the onset of pairing correlations, which translates into a rapid increase of the halo parameter. As was pointed out in \cite{HaloDuguet}, a shortcoming of the Helm method is that in some cases, the quantity $\delta R_{halo}$ is non-zero even in the middle of the valley of stability. This appears clearly in Fig. \ref{fig02} for the protons along the entire isotopic chain.

\begin{figure}[h]
\includegraphics[height=7.0cm,width=9.0cm]{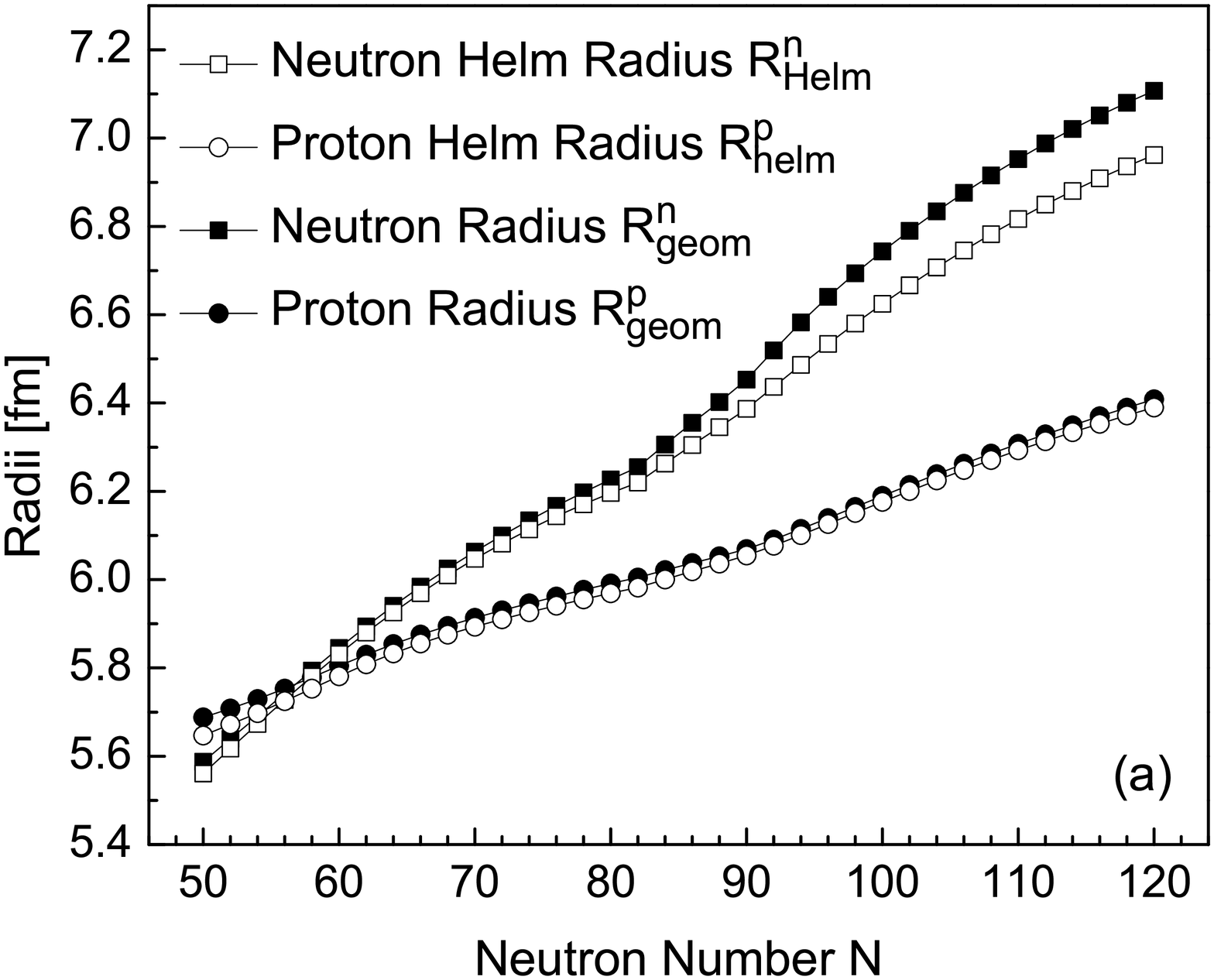}
\includegraphics[height=7.0cm,width=9.0cm]{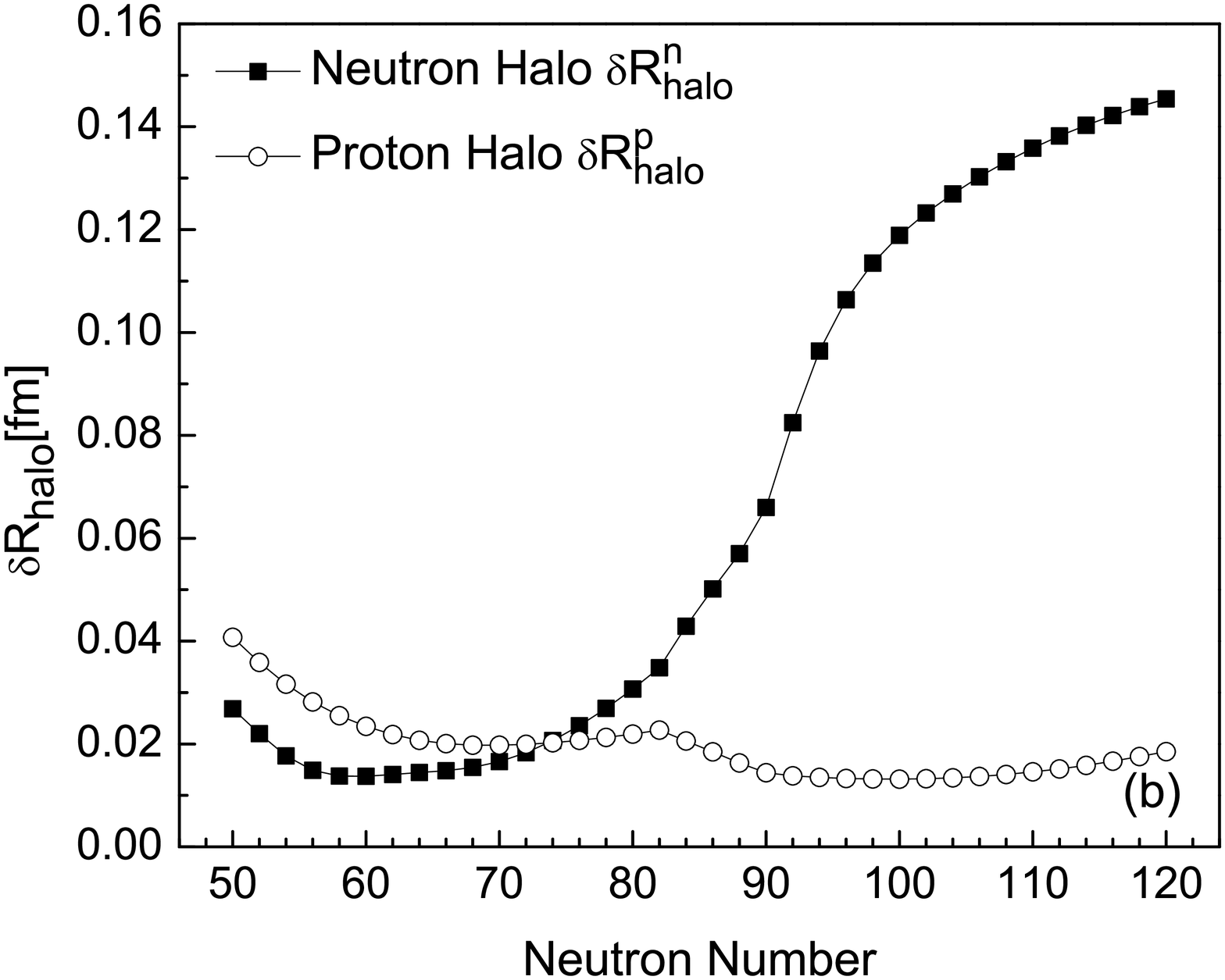}
\caption{Upper panel: Neutron $R_{geom}(n)$ and $R_{Helm}(n)$ and proton $R_{geom}(p)$ and $R_{Helm}(p)$ radii for the Sn isotopes calculated with the D1S Gogny interaction. Lower panel: Neutron and proton halo parameter $\delta R_{halo}$ along this isotopic chain.
		}
\label{fig03}	
\end{figure}

Another important remark is that both parametrizations of the Gogny force tend to give "compact" nuclei, with relatively small halos in agreement with those obtained  in \cite{HelmDoba} with the Skyrme SkP and relativistic mean-field NLSH and NL3 parametrizations and in contrast to the Skyrme/SLy4 interaction.
The case of Tin isotopes is even more enlightening. For this particular element, the position of the drip line is nearly identical in spherical HFB calculations with Gogny/D1S and Skyrme/SLy4 interactions, which facilitates the comparison. In the upper panel of Fig. \ref{fig03} we plot the neutron and proton geometrical and Helm radius from the proton to the neutron drip line. Note that near the neutron drip line, the halo is only about 0.15 fm while Skryme/SLy4 results reported in \cite{HelmDoba} indicate a size of about 0.8 fm. 

It could also be tempting to apply our method in some of the {\it experimental} cases of nuclear halos. However, as hinted in the introduction, we are faced with one major difficulty: most of the halo candidates are very light nuclei with $Z\leq 6$ at the drip line like $^{11}$Li, $^{14}$Be and $^{19}$B. For all the elements with $Z\leq 8$ the experimental drip line is rigorously established, in the sense that isotopes beyond the drip line are proved to be particle-unstable \cite{exp_dripline}. The application of our spherical Gogny-HFB calculations, whether the D1 or D1S interaction is used, gives the correct drip line isotope for Lithium (see next section) but fails to reproduce the experimental data for elements B, C and O. Three main mechanisms, possibly combined, can be the source of this discrepancy: (i) the interactions used are not extrapolable in these light nuclei (ii) additional mean-field symmetries must be broken, e.g. rotational invariance (iii) correlations beyond the HFB level must be included. It is almost certain that the fit of the interactions can be improved, but it is today difficult to assess to which extent this would affect the predictions of nuclear halos in very light nuclei. Similarly, it is not very clear at the moment how halos are formed in deformed nuclei. 

\subsection{Discussion on giant halos}

Since our description contains the main ingredients for a proper description of the halo phenomenon, namely, a good pairing force, the incorporation of the continuum and eventually particle number projection (in the VAP approach) indispensable in a weak pairing regime we can confront our model  with  recent spectacular predictions about the existence of several giant halos in light and medium-mass nuclei. In Neon isotopes, {\it spherical} coordinate-space Relativistic Hartree-Bogoliubov (RHB) calculations predicted that giant halos could develop for a number of neutrons around 30 \cite{PRL_RMF_1}. In our {\it spherical} Gogny-HFB calculations the drip line is positioned at $N = 20$ for both D1 and D1S interactions, which falls a bit short of the last known bound Neon isotope at $N=24$ \cite{MgNature} and references therein. When deformation is included in the calculation, the position of the (current) drip line shifts to $N=24$ \cite{Gogny-MassTable}. The very stretched drip line reported in \cite{PRL_RMF_1} is somewhat surprising since the pairing channel was treated by using the D1S finite-range interaction. Moreover, up to $N=20$, RHB results for the r.m.s. radii are very similar to ours: for example at $N=20$ we find a neutron r.m.s radius of $r_{n} = 3.39$ fm with the D1S interaction, to compare with RHB results of $r_{n} \approx 3.42$ fm. 

\begin{figure}[h]
\includegraphics[height=7.0cm,width=9.30cm]{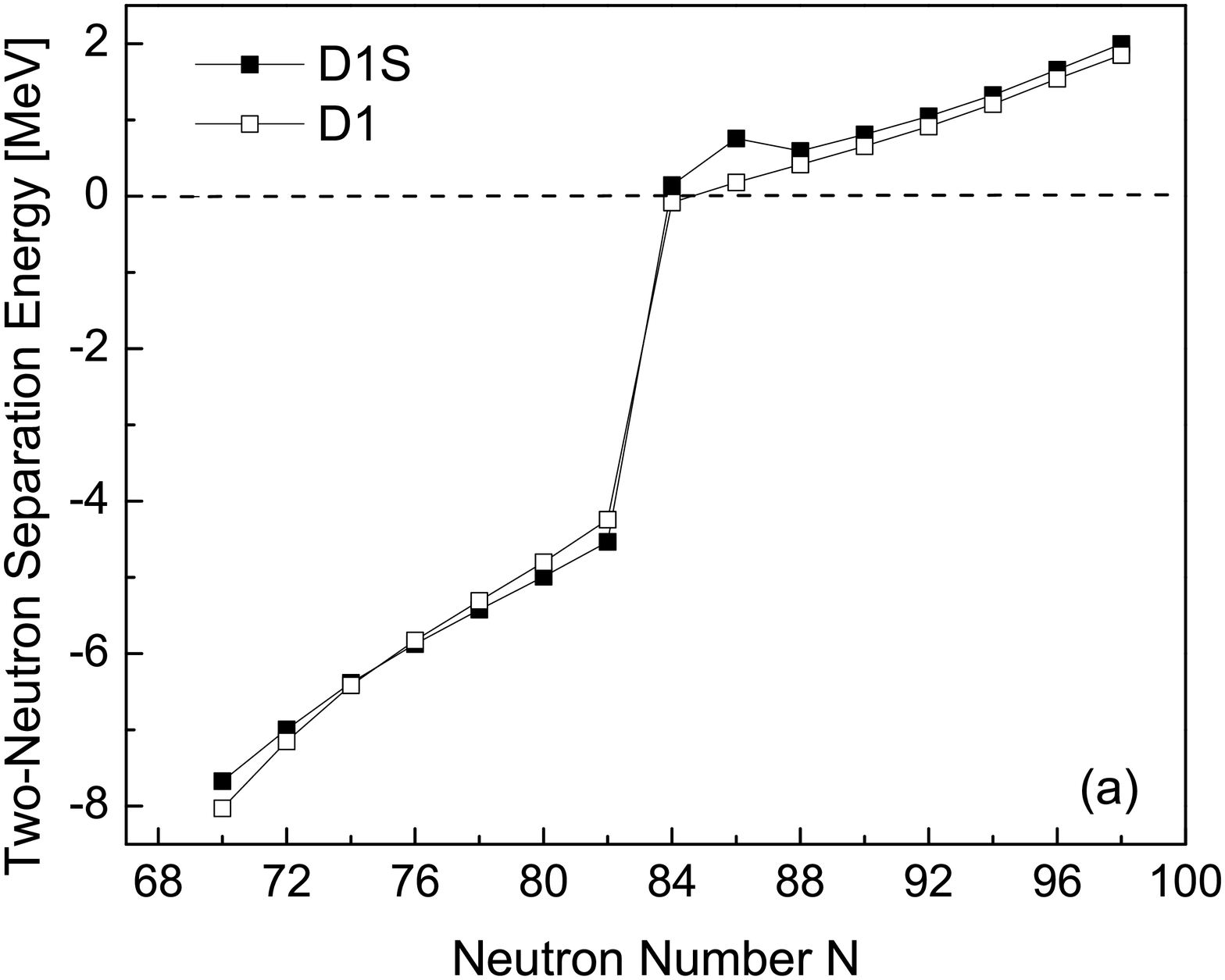}
\includegraphics[height=7.0cm,width=9.50cm]{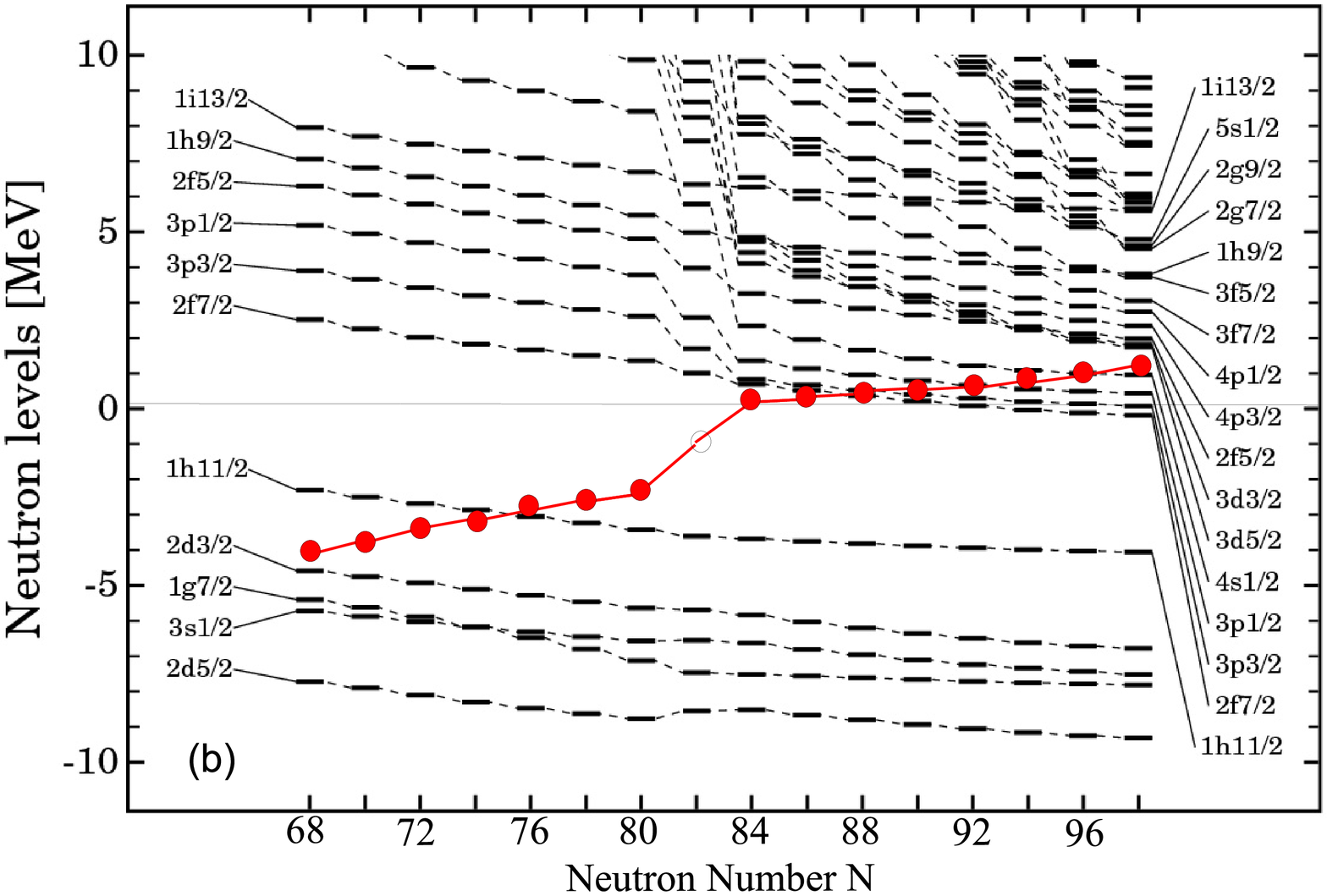}
\caption{(color on-line) Upper panel: Two-neutron separation energy for the Zr isotopic chain with the D1 and D1S parametrization. Lower panel: Neutron single-particle levels in the canonical basis for Zr isotopes (D1S interaction). The bullets represent the position of the Fermi level.}
\label{fig04}	
\end{figure}

\begin{figure}[t]
\includegraphics[height=6.5cm,width=9.0cm]{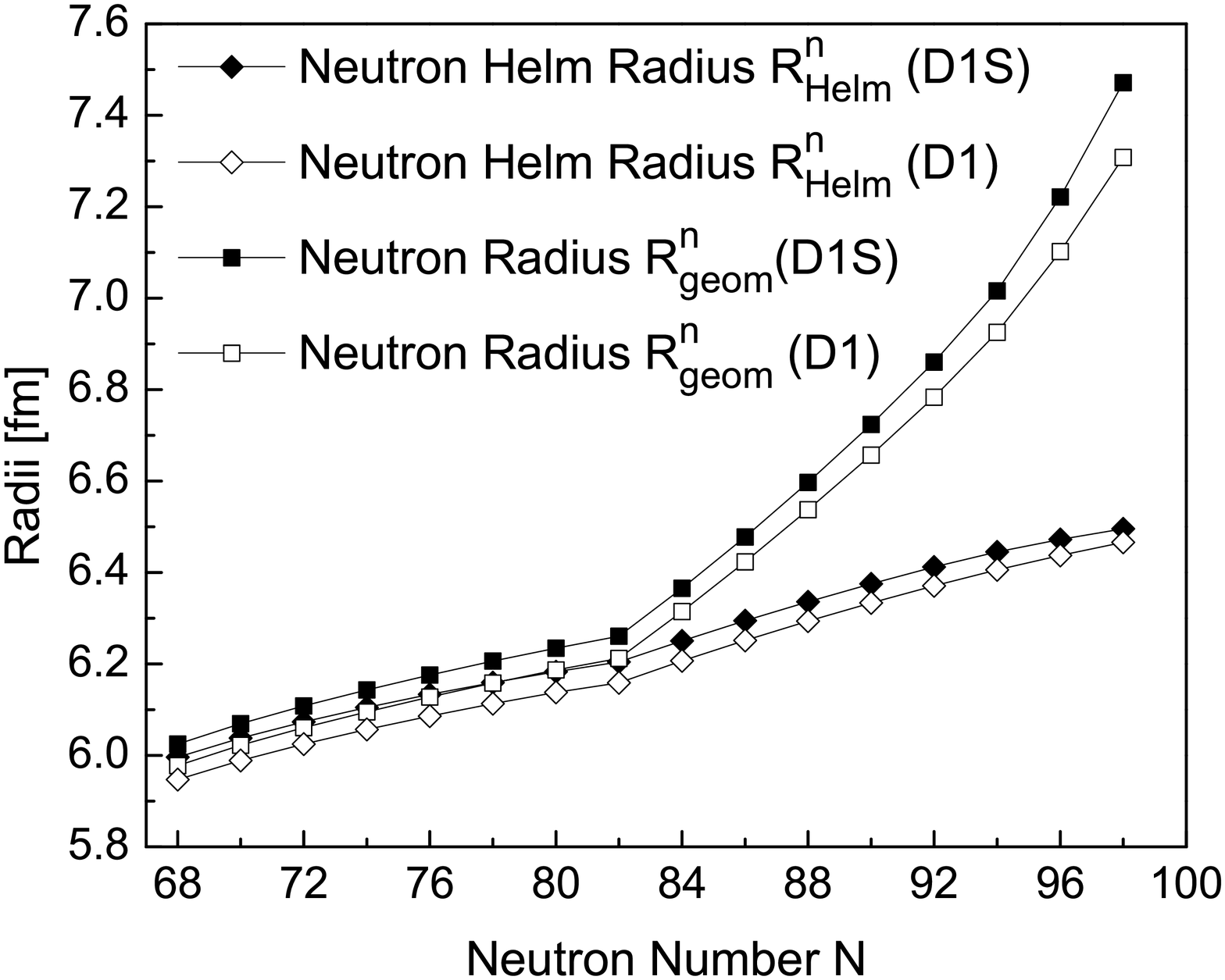}
\caption{ Neutron $R_{geom}(n)$ and $R_{Helm}(n)$ radii for the Zr isotopes calculated with the D1S (plain symbols) and D1 (open symbols) Gogny interaction. 
         }
\label{fig05}	
\end{figure}

The application of the spherical coordinate-space RHB, with a zero-range density-dependent force in the particle-particle channel, led to another prediction of giant halos in Zirconium isotopes \cite{PRL_RMF_2}. Similarly as in the case of Neon isotopes, such predictions are rooted in the existence of a very stretched drip line at $N = 100$ corresponding to element $^{140}$Zr. Results showed in Table \ref{table01} and upper panel of Fig.~\ref{fig04} show that the drip line in spherical Gogny-HFB calculations is at $N=82$ for the D1S and $N=84$ for the D1 interaction. Deformed Gogny-HFB calculations with the D1S interaction also suggest a drip line at $N=82$ \cite{Gogny-MassTable}. These results are in agreement e.g. with Skyrme HFB calculations with the SLy4 interaction \cite{stoitsov_dripline} which predict the drip line at $N=84$. Other parametrizations of the Skyrme interaction have slightly more extended drip lines, at N = 92 for SKP and N = 94 for SKM* \cite{stoitsov-masstable}. For the Gogny interaction, isotopes with $N\geq 82$ (D1S) and $N\geq 84$ (D1) are unbound with respect to two neutron emission, see upper panel of Fig. \ref{fig04}. Beyond drip line HFB calculations, although not realistic, can be pedagogical: in Fig. \ref{fig05} we show the evolution of the neutron geometrical and Helm radius beyond the drip line for both the D1 and D1S interactions. As we increase the number of neutrons, delocalized orbitals corresponding to discretized continuum states become occupied and cause a very marked increase of the neutron radius. 

If we restrict ourselves to physical solutions at the HFB level we find very small halos: $\delta R_{halo} \approx 0.06$ fm for the D1S interaction in $^{122}$Zr and $\delta R_{halo} \approx 0.11$ fm for the D1 interaction in $^{124}$Zr. One may be tempted to attribute this small value to the collapse of pairing correlations which occurs at $N = 82$. This collapse of pairing correlation can be inferred from the lower panel of Fig. \ref{fig04}, where the gap between the (occupied) h$_{11/2}$ and (empty) 2f$_{7/2}$ orbital is very large. However, particle-number projection before variation applied to this nucleus provides the same drip line Zr isotope and leads essentially to the same value of $\delta R_{halo}$ even though pairing correlations do not vanish any more. 

It should be noted that our results agree with previous works from \cite{PRL_RMF_2} (RHB), \cite{stoitsov-masstable} (SLY4, SKM* and SKP) as far as the main features of the shell structure of Zr isotopes are concerned, cf. for example the neutron single-particle levels in the canonical basis, Fig. 1 in \cite{PRL_RMF_2} and Fig. \ref{fig04} in the present work. In particular, the inflection point in the neutron radius at $N = 82$ is reproduced by all models. However, all three realizations of the nuclear mean-field differ as to the exact location of the drip line for Zirconium element, and it is this uncertainty that causes the widely different predictions of halo sizes in this particular element.

\begin{figure}[h]
\includegraphics[height=6.5cm,width=9.0cm]{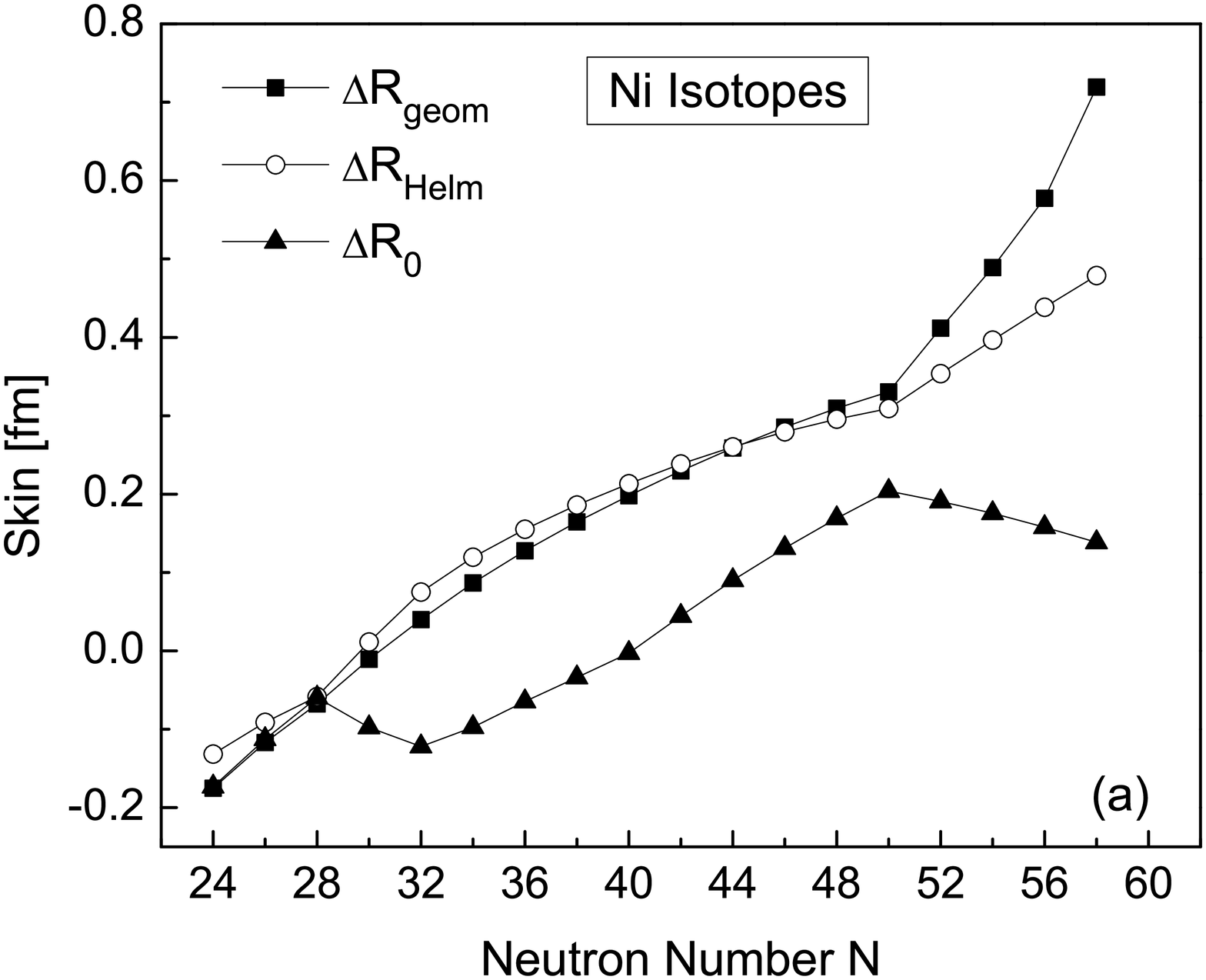}
\includegraphics[height=6.5cm,width=9.0cm]{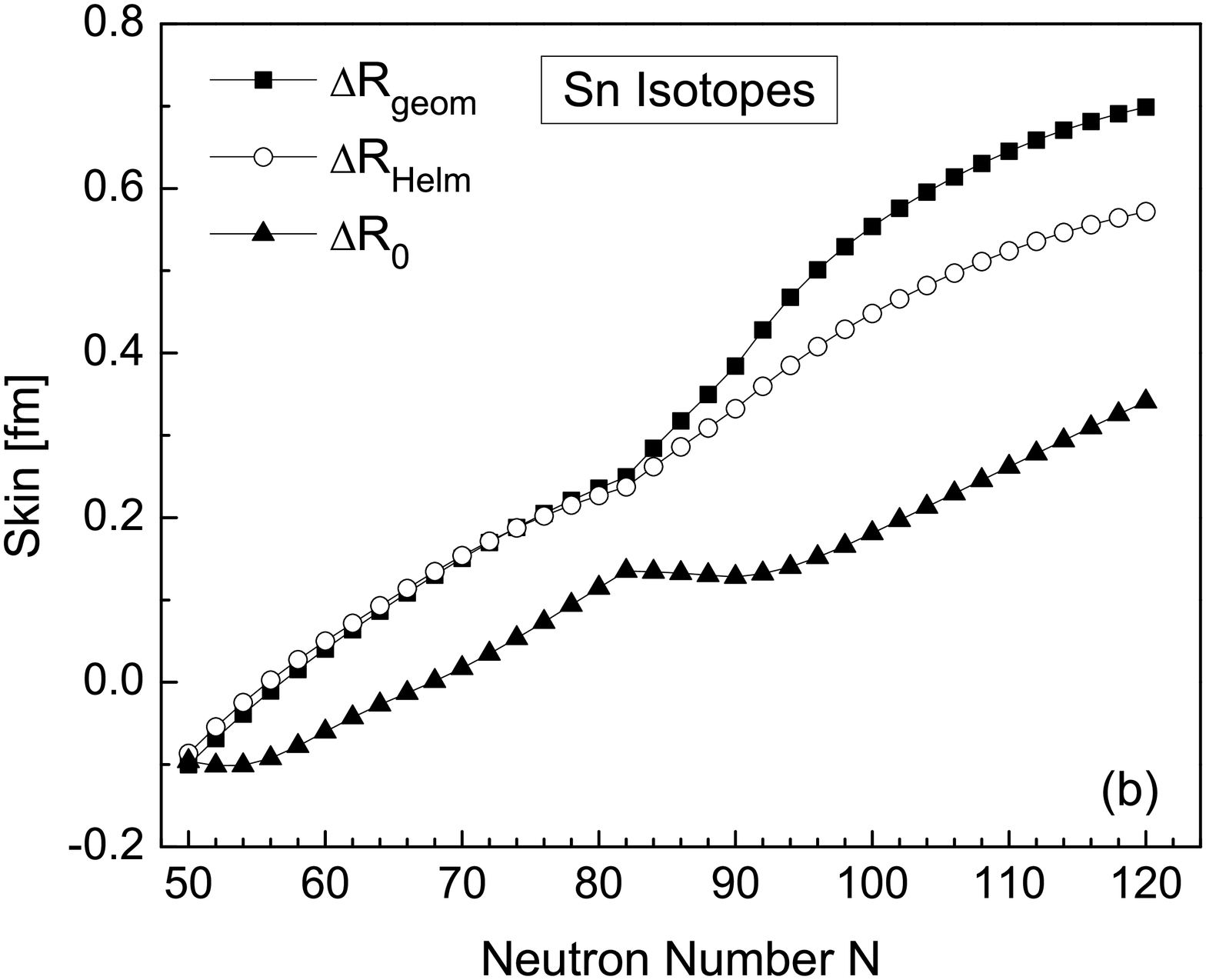}
\caption{Upper panel: Neutron skins for the Ni isotopes calculated with the D1S Gogny interaction and expressed as the difference of geometrical radii (plain squares), Helm radii (open circles) and diffraction radii (plain triangles). Lower Panel: Same figure as the upper panel for Sn element.
		}
\label{skins_Sn}	
\end{figure}

\subsection{Neutron skins}

One of the main features of neutron-rich nuclei is the development of neutron skins as the asymmetry between the number of neutrons and protons increases. The method that we developed to include the continuum in our calculations allow us to compute neutron skins up to the drip line. As an illustration, we display in Fig. (\ref{skins_Sn}) the neutron skins calculated from the geometrical, Helm and diffraction radius for the two isotopic chains of Nickel and Tin. 

As expected all three definitions of the neutron skin give a smooth increase with the neutron number. As noticed in \cite{HelmDoba}, however, the skin calculated from geometrical radii shows a clear inflection point at $N=50$ (Ni isotopes) and $N=82$ (Sn isotopes), which is directly related to the one marking the appearance of the neutron halo, cf.  Fig. \ref{fig03}. By contrast, the neutron skin calculated from the Helm radius is a more regular function of the neutron number. Interestingly, although the size of the halo with our Gogny/D1S interaction is markedly smaller than with e.g. Skyrme/SLy4, the values for the neutron skin are much closer: In $^{170}$Sn, $\Delta R_{Helm} \approx 0.57$ fm for D1S and $\Delta R_{Helm} \approx 0.70$ fm for Skyrme/SLy4 (a similar number is also obtained in Skyrme/SKP), cf. \cite{HelmDoba}.
    
\begin{figure}[h]
\includegraphics[height=7.0cm,width=9.50cm]{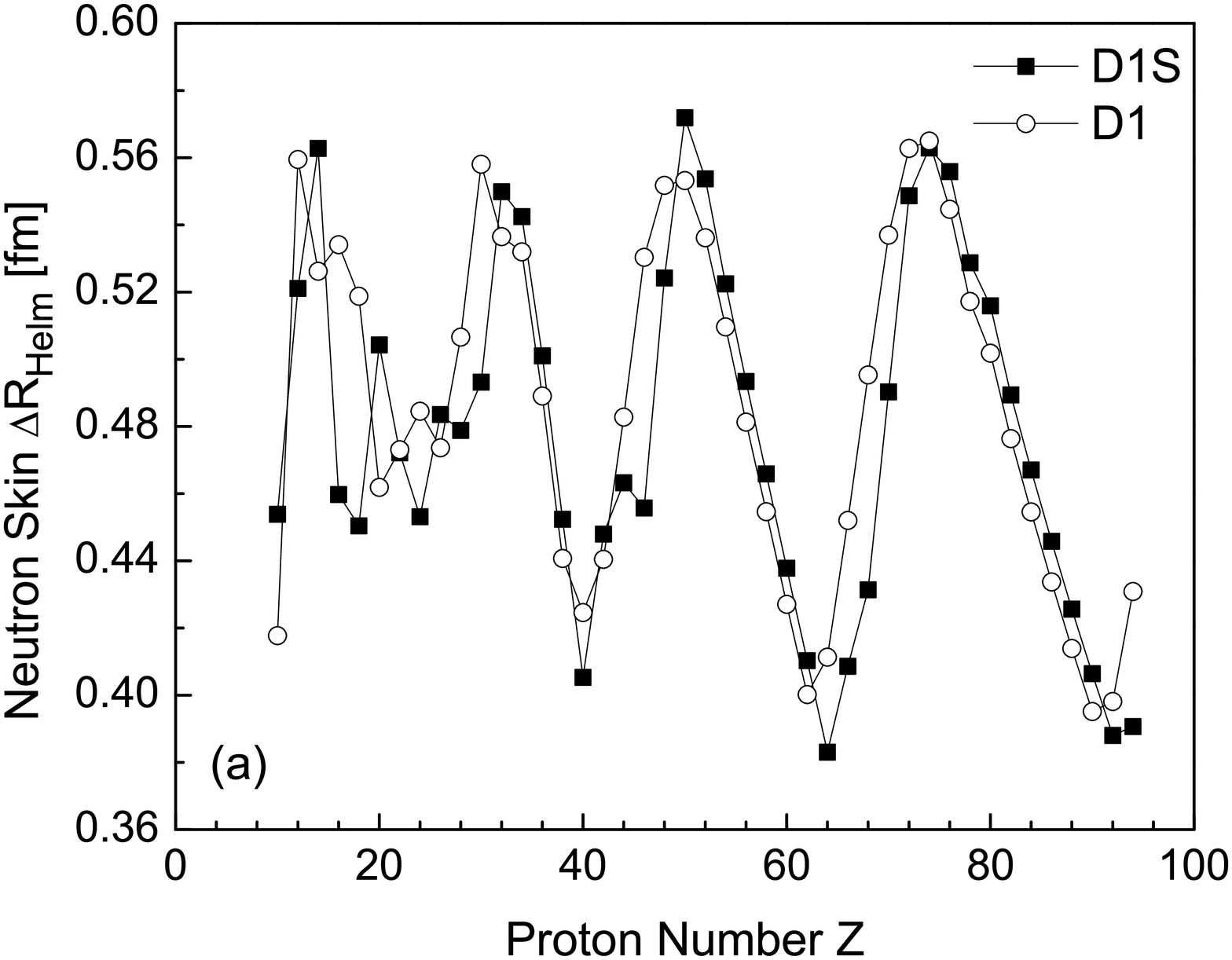}
\includegraphics[height=7.0cm,width=9.50cm]{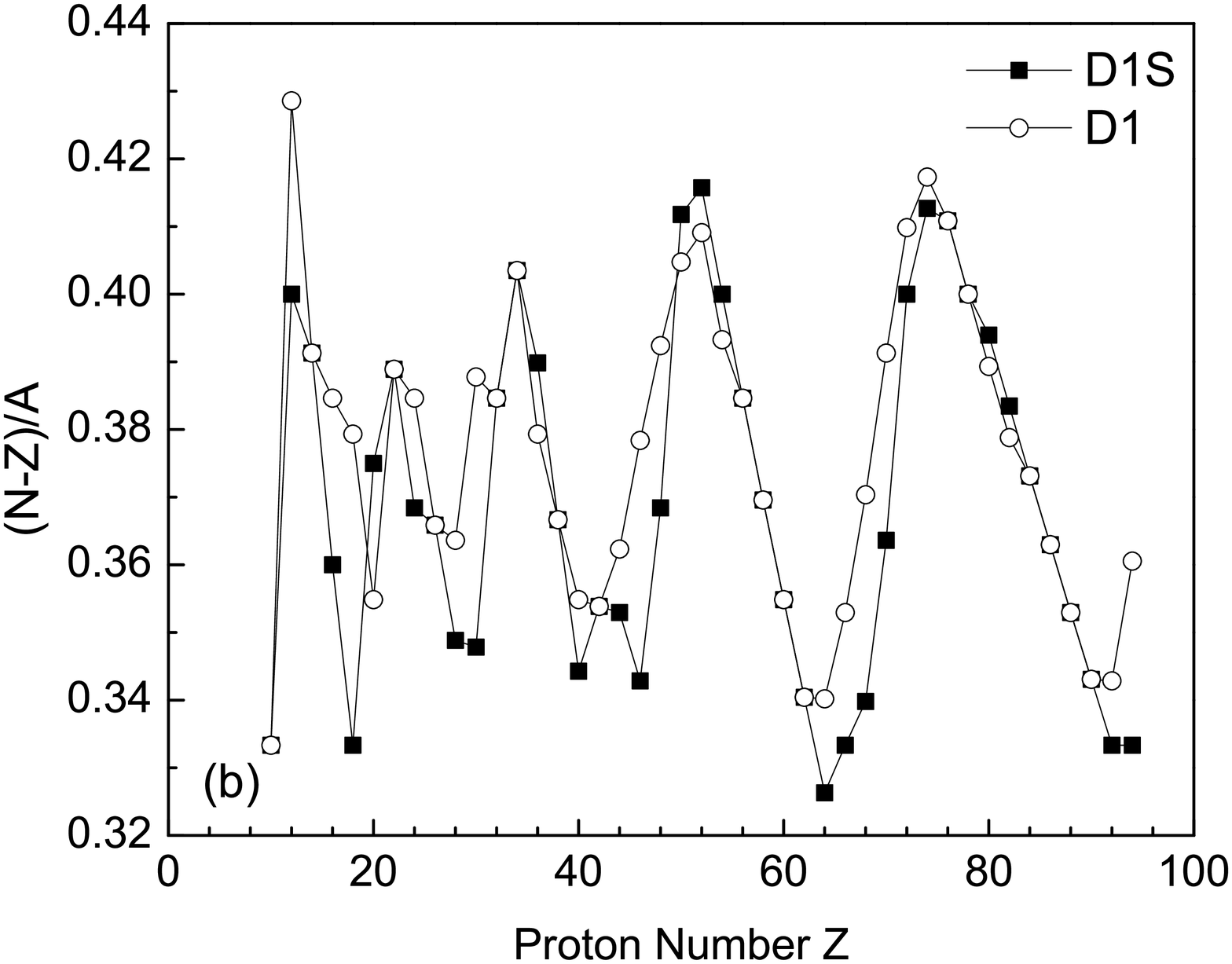}
\caption{Upper panel: Neutron skins along the neutron drip line calculated with the Gogny interaction and expressed as the difference of Helm radii with the D1S (plain squares) and D1 (open circles) interactions. Lower panel: Quantity $(N-Z)/A$ at the drip line for the D1S (plain squares) and D1 (open circles) interactions
         }
\label{neut_skins}	
\end{figure}

It is instructive to compute the neutron skin for all the elements located at the drip line. In the upper panel of Fig.(\ref{neut_skins}) we plot the neutron skin for the nuclei listed in Table \ref{table01}. We find an oscillatory behavior relatively similar to the one found for the quantity $\delta R_{halo}$ plotted in Fig. \ref{fig01}. In both cases, halos and skins, these oscillations can be somewhat related to neutron magic numbers, but the underlying physics is quite different. 

Skins are defined as the difference between the neutron and proton radius. Therefore, variations in the  shape  of the skins measures the relative increase or decrease of neutrons versus protons. At a neutron shell closure, one can add several protons without changing the position of the neutron drip line, i.e., as a function of Z, the proton radius increase and the neutron one remains constant, which produces a decrease in the neutron skin. This effect is very clearly seen in Fig. \ref{neut_skins}: the minima at $Z = 40$, $Z = 62-64$ (D1-D1S) and $Z = 90-92$ (D1-D1S) correspond to the last isotone with (magic) neutron number $N = 82$, $N = 126$ and $N=184$ respectively. Once beyond the neutron magic number, the neutron radius increases very rapidly, and this translates into a quick increase of the neutron skin for the next few elements. This sharp rise is also visible in Fig. \ref{neut_skins} in the range $40 \leq Z \leq 50$, $62-64 \leq Z \leq 72$.

In fact, the oscillations of the neutron skin can be correlated very neatly to the quantity (N-Z)/A. While the neutron excess N-Z increases with the mass number A (an upward trend not observed in Fig. \ref{neut_skins}), the ratio (N-Z)/A fluctuates around some average value of 0.37. In the lower panel of Fig.(\ref{neut_skins}) we plot (N-Z)/A  as a function of Z at the drip line for the two interactions D1S and D1 considered in this study. We observe that the maxima and minima of the neutron skin correspond almost exactly to the maxima and minima of (N-Z)/A, especially for heavy nuclei. In light nuclei, this correspondence remains, although it is a little less obvious. We should like to stress that the quantity (N-Z)/A is a direct measure of the ratio between the iso-vector and iso-scalar (integrated) densities. Neutron skins could therefore prove particularly useful to obtain experimental constraints on the corresponding terms of the interaction/functional.

It also follows from this observation that we do not observe for the neutron skins the clear downward trend as function of Z that was observed for the halos, cf. Fig. \ref{fig01}. Neutron skins are rather a mass independent observable, which implies that the skin in a very light nucleus such as, e.g. Si ($Z=14$) is of comparable size as the skin in W ($Z=74$). The amplitude of the oscillations is also much smaller for the skins than for the halos, reflecting the fact that the (N-Z)/A ratio does not vary too much along the neutron drip line. Also, the main maxima for halos and skins do not exactly coincide: For the D1S interaction, for example, the skins peak at $Z=14, 32, 50$ and 74 and the halos at $Z=16, 28, 44$ and 74.

\section{INFLUENCE OF SYMMETRY-RESTORATION ON NUCLEAR HALOS}

In this section, we discuss another mechanism that can affect the position of the drip line, namely the restoration of broken symmetries. We focus on the projection on good particle number before variation and examine several of its conceptual as well as practical consequences. The method we use to include continuum effects into our description of weakly-bound nuclei is indeed particularly suitable to include extensions beyond the mean-field. 

\subsection{The RVAP approach}

In \cite{nous} we briefly described how we can simulate the Variation After Projection (VAP) of the HFB solutions by means of the restricted-VAP (RVAP) method. Since along the drip lines some conceptual difficulties may arise we discuss the method a bit more at length. To illustrate how the RVAP method works we shall assume a generic two body Hamiltonian
\begin{eqnarray}
\hat{H}
=
\sum_{lq}t_{lq}c_{l}^{\dagger}c_{q}+\frac{1}{4}\sum_{lql'q'}\bar{v}_{lql'q'}c_{l}^{\dagger}c_{q}^{\dagger}c_{q'}c_{l'}
\label{hamil_c}
\end{eqnarray}
with $\bar{v}_{lql'q'}$ the antisymmetric matrix element 
\begin{eqnarray}
\bar{v}_{lql'q'}=v_{lql'q'}-v_{lqq'l'}
\label{potencialant}
\end{eqnarray}
and $(c_{i}^{\dagger},c_{i})$ the single-particle creation and annihilation operators in a given basis. Given the most general Bogoliubov transformation
\begin{equation}
\beta_{k}^{\dagger} = \sum_{l}U_{lk}c_{l}^{\dagger}+V_{lk}c_{l}
\label{betas}
\end{equation}
the HFB method provides the product wave function 
\begin{equation}
|\Phi\rangle=\prod_{q}\beta_{q}|-\rangle 
\end{equation}
that minimizes the expectation value of the Hamiltonian $\hat{H}$. The matrices $U$ and $V$ that fix the Bogoliubov transformation of Eq. (\ref{betas}) are determined by minimization of the functional
\begin{equation}
E_{HFB}'\left[|\Phi\rangle\right]=\frac{\langle\Phi|\hat{H}-\lambda_{N}\hat{N}-\lambda_{Z}\hat{Z}|\Phi\rangle}{\langle\Phi|\Phi\rangle}
\label{funcE}
\end{equation}
with $\lambda_N$ and $\lambda_Z$ the Lagrange parameters that adjust the average number of neutron and proton.

It can be shown \cite{RingSchuck} that the minimization of Eq.~(\ref{funcE}) amounts to the diagonalization of the matrix 
\begin{eqnarray}
\left(
\begin{array}{cc}
 h' & \Delta \\
 -\Delta^{*} & -h'^{*} \\  
\end{array}
\right)\left(
\begin{array}{c}
U_{k}\\
V_{k}\\  
\end{array}
\right) & = & E_{k} \left(
\begin{array}{c}
U_{k}\\
V_{k}\\  
\end{array}
\right) 
\label{hfb_eq_diag}
\end{eqnarray}
with $E_{k}$ the quasi-particle energies and  $h'=t+\Gamma-\lambda_{N}-\lambda_{Z}$. The Hartree-Fock field $\Gamma$ and the pairing field $\Delta$ are given by 
\begin{eqnarray}
\Gamma_{ll'}&=&\sum_{qq'}\bar{v}_{lql'q'}\rho_{q'q} \label{Gamma}\\
\Delta_{ll'}&=&\frac{1}{2}\sum_{qq'}\bar{v}_{ll'qq'}\kappa_{qq'} \label{Delta}
\end{eqnarray}
with $\rho$ the density matrix and $\kappa$ the pairing tensor defined by
\begin{eqnarray}
\rho_{ll'} & = & \langle\Phi|c_{l'}^{\dagger}c_{l}|\Phi\rangle=\left(V^{*}V^{T}\right)_{ll'} \nonumber\\
\kappa_{ll'} & = & \langle\Phi|c_{l'}c_{l}|\Phi\rangle=\left(V^{*}U^{T}\right)_{ll'}.
\end{eqnarray}

The particle number projected energy is given by
\begin{eqnarray}
E^{N}\left[|\Phi\rangle\right]
=
\frac{\langle\Phi^{N}|\hat{H}|\Phi^{N}\rangle}{\langle\Phi^{N}|\Phi^{N}\rangle} 
=
\frac{\langle\Phi |\hat{H}\hat{P}^{N}|\Phi\rangle}{\langle\Phi|\hat{P}^{N}|\Phi\rangle} 
\label{E_N}
\end{eqnarray}
with $\hat{P}^{N}$ the particle number projector and
\begin{equation}
|\Phi^{N}\rangle = \hat{P}^{N}|\Phi\rangle.
\label{WF_2}
\end{equation}
To avoid cumbersome formula we do not distinguish in Eq. (\ref{WF_2}) between protons and neutrons. The simplicity of projection techniques lies in the fact that, while $|\Phi^{N}\rangle$ is a correlated many-body wave function, the intrinsic wave function $|\Phi\rangle$ remains a product wave function, i.e. the variational parameters to be determined are the matrices $U$ and $V$ of Eq.~(\ref{betas}).

In the Variation After Projection (VAP) approach the projected energy $E^{N}$, see Eq.~(\ref{E_N}), is minimized directly. In the Projection After Variation (PAV) approach the HFB energy $E_{HFB}'$, see Eq.~(\ref{funcE}), is minimized first and the projection is carried out on the HFB wave-function after convergence. The difference is clear: in the VAP method we minimize the energy of the one nucleus (Z,N) we are interested in while in the PAV, the energy of a superposition of nuclei with numbers of particle Z and N around the actual values. Though the variational parameters are the same, the solution of the VAP equations is numerically much more involved than the PAV one. In a strong pairing regime the PAV solution might be a good approximation but in the general case, and in particular along the drip lines, the VAP one is much better. With finite range forces the solution of the VAP equations is rather involved, see \cite{PNP-Gogny}. Considering the additional difficulties inherent to a proper treatment of the coupling to the continuum, it is clear that a full VAP solution is beyond the actual numerical capabilities.

A way out of this problem is the Restricted VAP. In the VAP method the whole Hilbert space associated to the transformation Eq.~(\ref{betas}) is scanned in the variational procedure. In the RVAP approach, however, only a restricted variational space of highly correlated wave-functions is allowed. In our case, since we are interested in pairing correlations, our restricted space should contain a whole set of paired wave-functions $|\Phi(\delta)\rangle$ which parametrically depend on the real number $\delta$. To generate such wave-functions with different pairing content and  that  simultaneously are consistent with our Hamiltonian, we proceed in the following way: Instead of iterating Eq.~\ref{hfb_eq_diag} together with Eqs.~(\ref{Gamma},\ref{Delta}) as in the usual HFB case, we now iterate 
\begin{eqnarray}
\left(
\begin{array}{cc}
 h' & \delta \cdot \Delta \\
 -\delta \cdot \Delta^{*} & -h'^{*} \\  
\end{array}
\right)\left(
\begin{array}{c}
U_{k}\\
V_{k}\\  
\end{array}
\right) & = & E_{k} \left(
\begin{array}{c}
U_{k}\\
V_{k}\\  
\end{array}
\right) 
\label{hfb_eq_diagm}
\end{eqnarray}
together with Eqs.~(\ref{Gamma},\ref{Delta}) until the convergence is achieved. The matrices $U(\delta)$ and $V(\delta)$ obtained in this way determine the wave-functions $|\Phi(\delta)\rangle$. Performing the same procedure for different $\delta$ values we generate the restricted correlated Hilbert space. We then project these wave-functions onto good-particle number and obtain a family of particle-number projected states $|\Phi^N(\delta)\rangle = \hat{P}^{N}|\Phi(\delta)\rangle$, for $\delta = 1.0, \dots, \delta_{max}$. The range of values for $\delta$ is chosen in such a way that at least several $|\Phi^N(\delta)\rangle$ wave functions correspond to highly-paired states. We can then take the expectation value of the  Hamiltonian with this set of wave-functions, i.e, using eq.~(\ref{E_N}). This gives us a curve $E^{N}(\delta)$ where, at each point $\delta$, the particle number is conserved. The variational principle guarantees that such a curve has a minimum, which approaches the VAP result \cite{restricted-VAP} .
 
\begin{figure}[h]
\includegraphics[height=7.0cm,width=9.0cm]{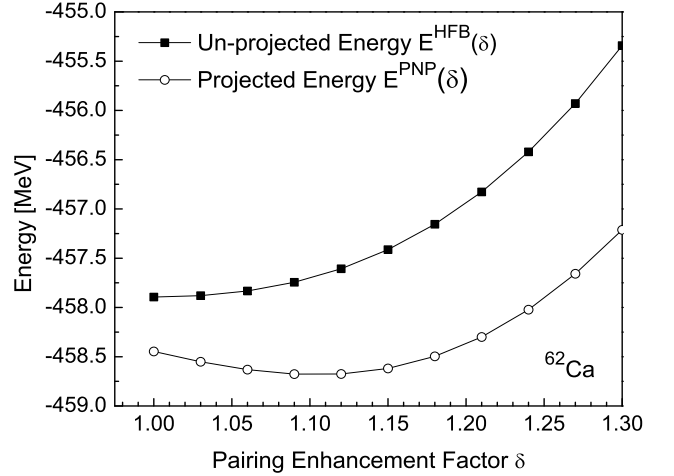}
\caption{Intrinsic (plain squares) and projected (open circles) energy as a function of the parameter $\delta$ for the nucleus $^{62}$Ca.}
\label{int_proj_ener}	
\end{figure}

To illustrate the procedure with a numerical application in Fig.~\ref{int_proj_ener} we display the un-projected energy
 \begin{equation}
E^{HFB}(\delta)= \frac {\langle \Phi(\delta )|\hat{H}|\Phi(\delta )\rangle}{\langle \Phi(\delta )|\Phi(\delta )\rangle}
 \end{equation}
 and the projected one 
  \begin{equation}
    E^{PNP}(\delta)=\frac{\langle \Phi(\delta )|\hat{H}\hat{P}^N|\Phi(\delta )\rangle}{\langle \Phi(\delta )|\hat{P}^{N}|\Phi(\delta )\rangle}
 \end{equation}
for the drip line nucleus $^{62}$Ca with the D1S interaction. When computing the density-dependent contribution to the projected energy $E^{PNP}(\delta)$, the projected density $\rho^{PNP}$ has been used (prescription 1 in \cite{PNP-Gogny}).

Since the HFB self-consistent minimum is obtained, by definition, at $\delta =1$ for $E^{HFB}(\delta)$ we expect a parabolic behavior around this value for increasing or decreasing $\delta$ values. Concerning  $E^{PNP}(\delta)$, at $\delta = 1$ projecting the HFB solution onto good particle number lowers the energy and for increasing pairing correlations, i.e., values of $\delta$ larger than 1, we first observe a decrease of the projected energy up to a minimum around $\delta =1.12$ followed by a rapid increase. Obviously the solution of the RVAP approach is $|\Phi(\delta=1.12)\rangle$.
 
In a PNP approach the drip lines are defined in terms of the projected separation energies, i.e. in terms of $S^N_n =B^{N}(N,Z)-B^{N-1}(N-1,Z)$ and $S^N_{2n} =B^{N}(N,Z)-B^{N-2}(N-2,Z)$. In the HFB approach $S_n\approx - \lambda_n$ and the one neutron drip line can be easily calculated. This approximation is not valid any more in a projected theory and $S^N_n$ must be explicitly calculated. Since for the moment we are not able to project on an odd number of particles we cannot calculate the one neutron drip line and will therefore focus on the 2-neutron drip line. In the next section we discuss the meaning of $\lambda$ and other quantities in the particular context of  a particle number projected theory.

\subsection{On the RVAP approach and the number of particles of the intrinsic wave function}

Let $|\Phi \rangle$ be a HFB wave-function, i.e. a particle number symmetry violating wave function. We will now show that the particle number projected energy is invariant under transformations that change the particle number of the underlying HFB wave function. We define
\begin{equation}
   |\tilde{\Phi} \rangle = e^{\alpha \Delta \hat{N}} | \Phi \rangle
\label{Transfo_N}   
\end{equation}  
with $\Delta \hat{N} = \hat{N} -N_0$,  $N_0= \langle \Phi |\hat{N} |\Phi\rangle$ and $\alpha$ is a real number and we assume that 
$\langle \Phi | \Phi\rangle=1$. The wave function $| \Phi \rangle$ can be written as \cite{RingSchuck} 
\begin{equation}
   |{\Phi} \rangle = \sum_{\beta^{\prime},N^{\prime}} C_{\beta^{\prime},N^{\prime}} |\beta^{\prime}, N^{\prime} \rangle
\end{equation}  
where $|\beta^{\prime}, N^{\prime} \rangle$ is an eigenstate of $\hat{N}$  with particle number $N'$ and $\beta'$ stands for all other necessary quantum numbers. The transformed wave function reads:
\begin{eqnarray}
|\tilde{\Phi} \rangle 
& =  &
e^{\alpha \Delta \hat{N}} \sum_{\beta^{\prime},N^{\prime}} C_{\beta^{\prime},N^{\prime}} |\beta^{\prime}, N^{\prime} \rangle \\
& = &
\sum_{\beta^{\prime},N^{\prime}} C_{\beta^{\prime},N^{\prime}} e^{\alpha  {(N^{\prime}-N_0)}}    |\beta^{\prime}, N^{\prime}\rangle 
\end{eqnarray}  
The projected energy is given by
\begin{eqnarray}
E^N
& = &
\frac{\langle \tilde{\Phi}|  \hat{H} \hat{P}^{N} |  \tilde{\Phi}\rangle}{\langle \tilde{\Phi}|  \hat{P}^{N} |  \tilde{\Phi}\rangle} \\
& = &
\frac{\sum_{\beta{\prime}{\prime},\beta^\prime} C_{\beta^{\prime \prime},N}^* C_{\beta^\prime,N}\langle \beta^{\prime\prime}N | \hat{H} |\beta^\prime N \rangle}
 {\sum_{\beta^{\prime \prime},\beta\prime} C_{\beta^{\prime \prime},N}^* C_{\beta^\prime,N}\langle \beta^{\prime \prime} N |\beta^\prime N \rangle} \\
& =&
\frac{\langle {\Phi}| \hat{H} \hat{P}^{N} |  {\Phi}\rangle}{\langle {\Phi}|  \hat{P}^{N} |  {\Phi}\rangle}
\end{eqnarray}  
in an obvious way.

The wave functions $ |\tilde{\Phi} \rangle$ and $ |{\Phi} \rangle$ have different number of particles on the average. This can be easily shown assuming the parameter $\alpha$ small enough, in this case
\begin{equation}
\frac{\langle \tilde{\Phi}| \hat{N} |  \tilde{\Phi}\rangle} {\langle \tilde{\Phi}| \tilde{\Phi}\rangle}
= 
N_0 + 2\alpha \langle \Phi | (\Delta \hat{N})^2 |  {\Phi}\rangle.
\label{end_Transfo_N}
\end{equation}  
up to $\alpha^2$ terms. Since $|\Phi \rangle$ is per definition a symmetry violating wave function, $\langle \Phi |(\Delta \hat{N})^2 |  {\Phi}\rangle\neq 0$, the wave functions $|\Phi \rangle$ and $|\tilde{\Phi} \rangle$ do have on the average different number of particles.

We have therefore demonstrated that we can change the average number of particles of the intrinsic wave function without changing the value of the projected energy: The Lagrange parameter $\lambda$ is therefore superfluous. As a matter of fact, in a VAP approach one uses a Lagrange parameter only to speed up the convergence of the iterative procedure.
 
It is important to realize that in a projected theory the only meaningful quantities are the projected ones. For example, the intrinsic density $\rho(\vec{r})$ is not invariant under the transformations of Eq. (\ref{Transfo_N}). This is simply the mathematical transcription of the fact that changing the particle number affects the intrinsic density. Conversely the projected density $\rho^N(\vec{r})$ is consistently invariant under the aforementioned transformation. 
   
In the demonstration above, Eqs.(\ref{Transfo_N})-(\ref{end_Transfo_N}) we have assumed that the coefficients of the Bogoliubov transformation (\ref{betas}) are known. In the case of the full VAP approach, this is automatically the case because the $U$ and $V$ matrices are determined self-consistently by minimizing the projected energy, which is invariant under transformations that change the number of particles. In the RVAP approach, however, to determine the $U(\delta)$ and $V(\delta)$ matrices one solves the standard HFB equations with a constraint on the number of particles. The latter equations are obviously not invariant under the transformations of Eq. (\ref{Transfo_N}). This may generate a dependence of the RVAP solution on the Lagrange parameter $\lambda$ (or equivalently on $\langle \hat{N} \rangle$). Obviously 
\begin{equation}
\lambda_n (\delta) = \frac{d\langle \Phi (\delta)| \hat{H} |\Phi (\delta)\rangle}{d\langle \Phi (\delta)| \hat{N} |\Phi (\delta)\rangle} 
\approx - S_n ^{HFB}  (\delta) \neq - S_n ^{PNP}  (\delta),
\end{equation}
which illustrates that $\lambda_n$ can not be used to define the one neutron drip line in a PNP approach. It is interesting  to realize that this dependence on $\lambda$ could be eventually used to generate additional correlated wave functions $|\Phi(\delta,\lambda)\rangle$ in the RVAP approach, thereby lowering further the projected energy \cite{FE.05}. 

\begin{figure}[h]
\includegraphics[height=7.0cm,width=9.0cm]{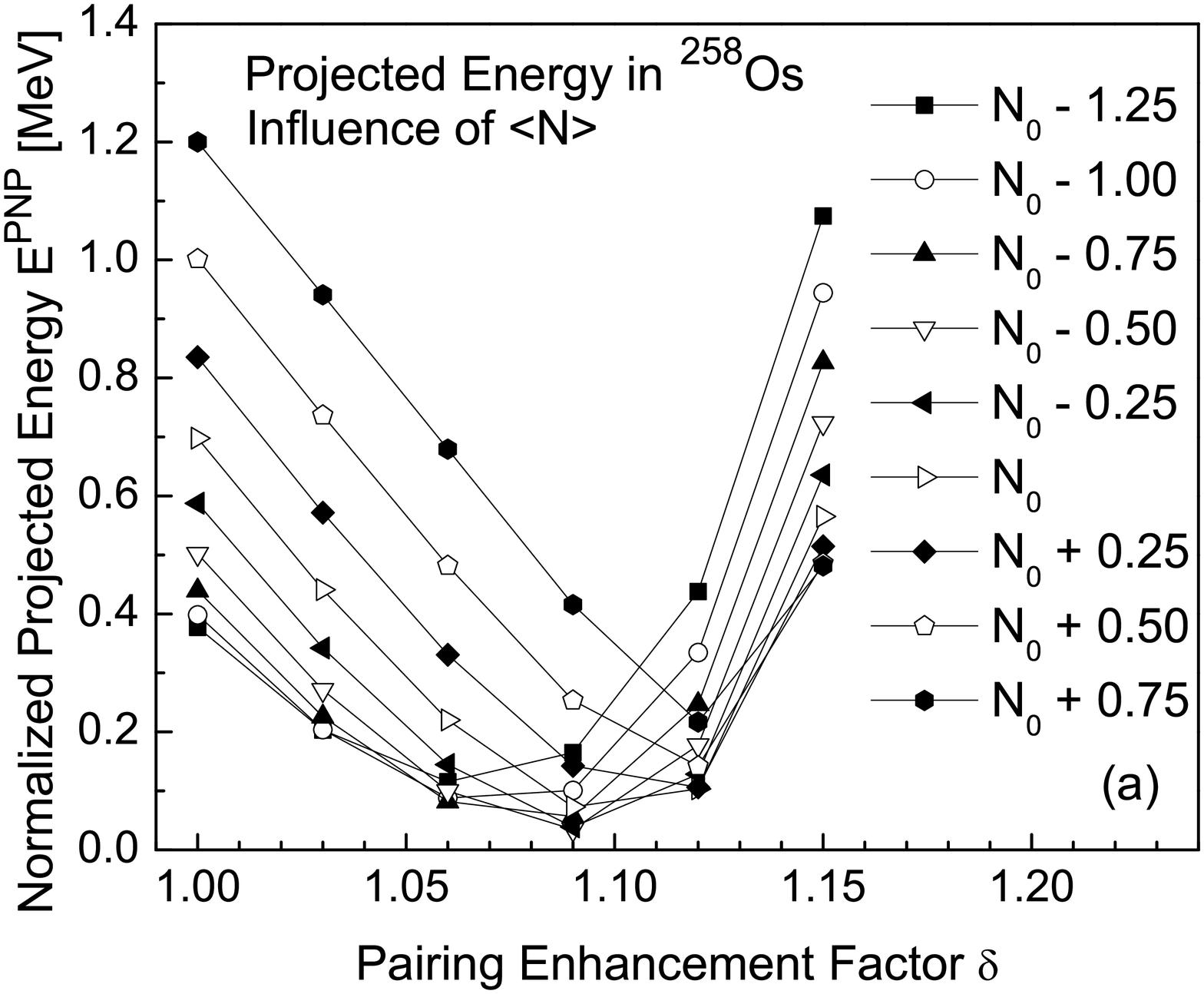}
\includegraphics[height=7.0cm,width=9.0cm]{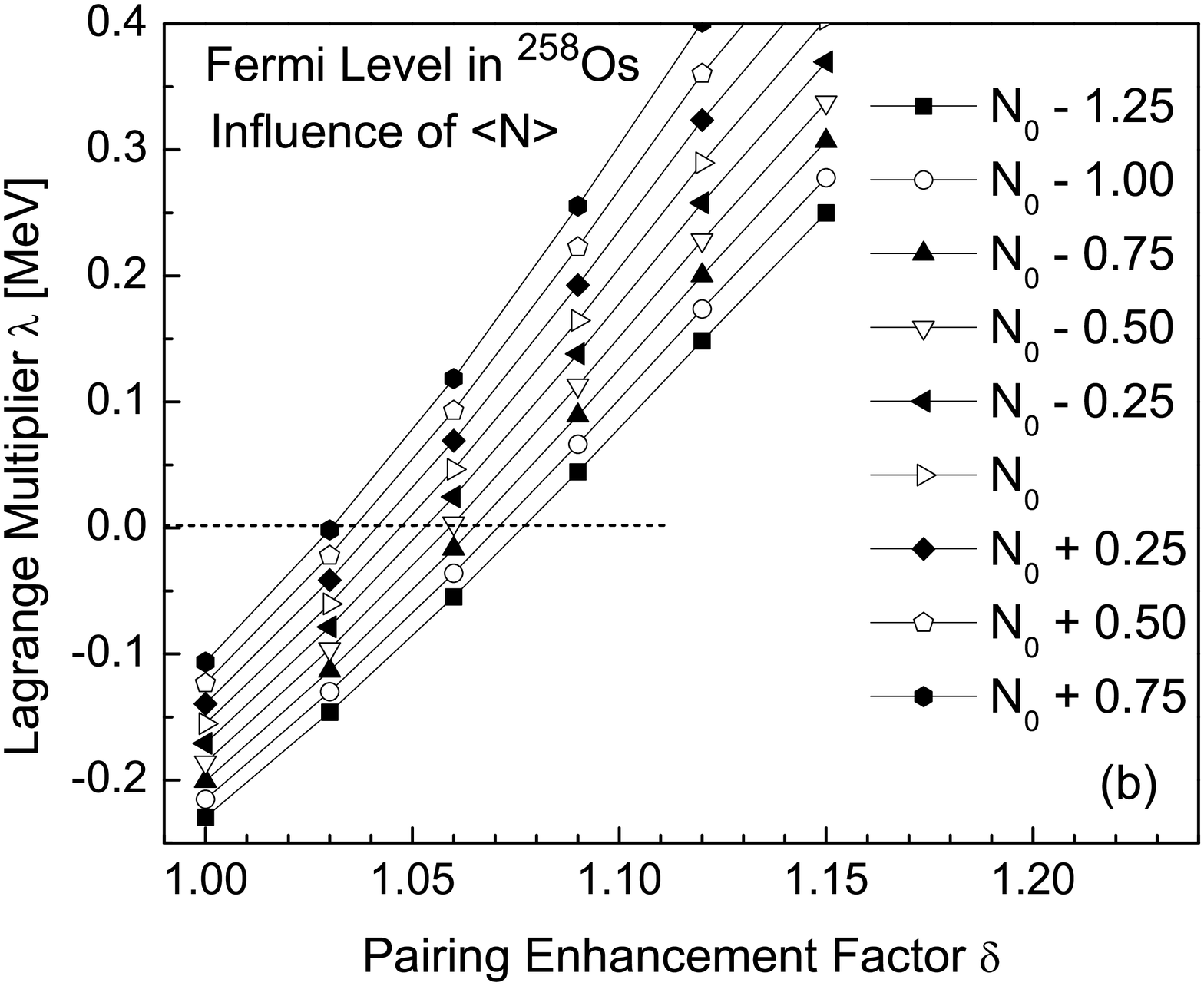}
\caption{Upper panel: Projected energy for intrinsic wave functions with different average number of particles $N$ around the actual particle number $N_{0} = 182$ in $^{258}$Os. This corresponds to intrinsic wave functions with different Lagrange multipliers. Lower panel: corresponding Fermi level $\lambda$.
	 }
\label{fig06}	
\end{figure}

As an illustration we show in the upper panel of Fig.~\ref{fig06} the projected energy of $^{258}$Os as a function of $\delta$ for different values of the number of particles (or $\lambda$) of the intrinsic wave function. As expected, we find that the minimum of the projected energy does not always correspond to the constraint $\langle N\rangle = N_0$, and for a given constraint on the average particle number, the position of the minimum depends on $\delta$. In the lower panel the corresponding chemical potentials $\lambda (\delta)$ are plotted. 

If one restricts oneself to "one-dimensional" RVAP wave functions of the type $|\Phi(\delta)\rangle$, it may happen, in particular near the drip lines, that in the RVAP minimum, the underlying HFB wave function $|\Phi(\delta)\rangle$ corresponds to a positive value of $\lambda$. As emphasized before, this is with no consequence since this $\lambda$ parameter does not define the drip line. If one insists, however, in having a negative Fermi energy, it is always possible to slightly change the average number of particles of the intrinsic wave function in such a way that $\lambda$ becomes negative, with the eventual cost of a small energy loss. In the illustrative case of $^{258}$Os displayed in Fig.~\ref{fig06} the energy cost is approximately 15 keV to go from the RVAP minimum (at $\delta = 1.09$) built on the HFB solution with average number of particles $\langle \hat{N} \rangle = N_{0}$, to the RVAP minimum (at $\delta = 1.06$) with $\langle \hat{N} \rangle = N_{0} - 1.0$. The Fermi energy of the underlying HFB solution goes from +165 keV to -36 keV. The drip line, defined from the two-neutron separation energy $S_{2n}$, remains unchanged.

\subsection{Nuclear halos and drip lines in a symmetry conserving approach}

In this section we investigate the effect of the particle number projection on the size of the halo along the neutron drip line. As mentioned in the Introduction we should distinguish between halos in very light nuclei and in the heavier ones. We are aware that a mean field based approach may not contain enough correlations to describe the halo mechanism in very light nuclei. Nevertheless we shall first discuss the impact of the RVAP procedure on the archetypical case of halo nucleus $^{11}$Li.

\begin{figure}[h]
\includegraphics[height=7.0cm,width=9.0cm]{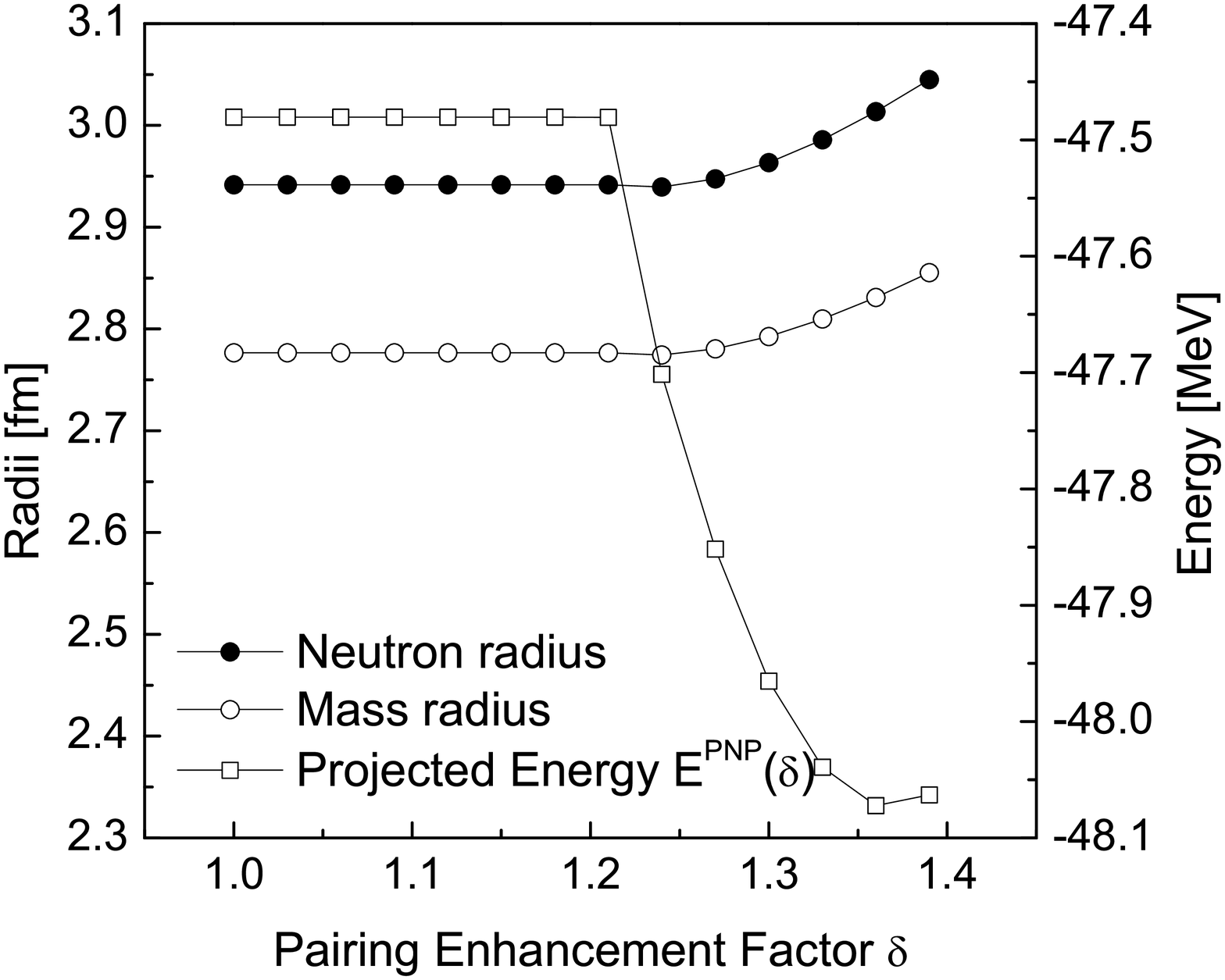}
\caption{Total projected energy (open squares) and neutron (plain circles) and mass (open triangles) r.m.s radius in $^{11}$Li as a function of the RVAP parameter $\delta$. At each point $\delta$ the HFB solution is projected on good particle number. The minimum is attained at $\delta=1.36$.
	 }
\label{fig07}	
\end{figure}

In the  calculation of the nucleus $^{11}$Li, with 3 protons and  8 neutrons, the odd proton was treated in the equal-filling approximation (the 1p$_{1/2}$ state is the blocked state) and only the projection on neutron particle number was carried out. Since the neutron number corresponds to a shell closure it is obvious that the HFB solution is not a super-fluid one. Figure \ref{fig07} shows the total projected energy and the neutron and mass r.m.s radius in $^{11}$Li as function of the RVAP variational parameter $\delta$. All calculations are done in the WS basis with the D1S interaction. The RVAP minimum always corresponds to a paired solution. In $^{11}$Li, our original spherical HFB calculations with the D1S or D1 interactions do not produce any halo. In fact, pairing correlations do not set in at all in this nucleus in the HFB calculations, even when the size of the box is increased up to 30 fm (thereby increasing the level density of continuum states). This is clearly viewed in Fig. \ref{fig07} since the projected energy remains constant at $E^{N} = -47.48 $ MeV for $1.0 \leq \delta \leq 1.18$. In spite of multiplying the pairing field by the factor $\delta$ during the iterations, pairing correlations are still identically 0 at convergence. Only for $\delta>1.18$ do we observe the onset of significant pairing correlations. The total projected energy therefore decreases, continuum states begin to have a non-zero occupation probability, which contributes to the increase of the r.m.s neutron radius. At the minimum of the RVAP curve, both the neutron and mass radius have increased by about 2 \%. The effect is marked but it is clearly not enough to reproduce the experimental halo in this nucleus \cite{halo_11Li}.

\begin{figure}[h]
\includegraphics[height=7.0cm,width=9.0cm]{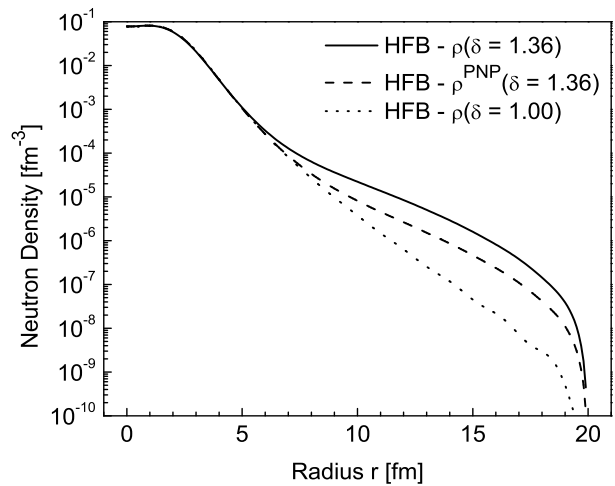}
\caption{Neutron HFB density $\rho^{HFB}(r)$ in $^{11}$Li at the HFB minimum (dotted line) and at the RVAP minimum (plain line). The dashed line shows the projected density $\rho^{PNP}(r)$ at the RVAP minimum. Calculations are done for the D1S interaction in the WS basis with $R_{box}=20$ fm.
	 }
\label{fig08}	
\end{figure}

To better grasp the impact of particle-number projection we show in Fig. \ref{fig08} the neutron density in $^{11}$Li in 3 different cases. The dotted line corresponds to the standard  HFB calculation. The plain line corresponds to the density of the intrinsic HFB wave function $|\Phi(\delta)\rangle$ at the RVAP minimum $\delta=1.36$. We clearly see the formation of a "bump" which is a visual trademark of the nuclear halo. However, this solution is not physical since it is only used to generate the variational space used in the RVAP procedure. Only the projected solution $|\Phi^{N}(\delta=\delta_{min})\rangle$ at the minimum is physical. The corresponding projected density (dashed line) is slightly less extended than the underlying HFB solution. 

\begin{table}[h]
\begin{center}
\caption{Table of  two neutron drip line nuclei obtained using the D1S parametrization at the HFB and RVAP level.}
\begin{ruledtabular}
\begin{tabular}{cc|cc}
   HFB  & RVAP &  HFB  & RVAP \\
\hline
  $^{~20}$C  &  $^{~20}$C  &  $^{170}$Sn &  $^{172}$Sn \\
  $^{~26}$O  &  $^{~26}$O  &  $^{178}$Te &  $^{178}$Te \\
  $^{~30}$Ne &  $^{~30}$Ne &  $^{180}$Xe &  $^{180}$Xe \\
  $^{~40}$Mg &  $^{~42}$Mg &  $^{182}$Ba &  $^{182}$Ba \\
  $^{~46}$Si &  $^{~46}$Si &  $^{184}$Ce &  $^{184}$Ce \\
  $^{~50}$S  &  $^{~50}$S  &  $^{186}$Nd &  $^{186}$Nd \\
  $^{~58}$Ar &  $^{~58}$Ar &  $^{188}$Sm &  $^{188}$Sm \\
  $^{~64}$Ca &  $^{~62}$Ca &  $^{192}$Gd &  $^{192}$Gd \\
  $^{~72}$Ti &  $^{~72}$Ti &  $^{200}$Dy &  $^{200}$Dy \\
  $^{~78}$Cr &  $^{~74}$Cr &  $^{206}$Er &  $^{206}$Er \\
  $^{~84}$Fe &  $^{~84}$Fe &  $^{220}$Yb &  $^{220}$Yb \\
  $^{~86}$Ni &  $^{~86}$Ni &  $^{242}$Hf &  $^{242}$Hf \\
  $^{~94}$Zn &  $^{~94}$Zn &  $^{254}$W  &  $^{254}$W  \\
  $^{104}$Ge &  $^{104}$Ge &  $^{258}$Os &  $^{258}$Os \\
  $^{114}$Se &  $^{114}$Se &  $^{260}$Pt &  $^{262}$Pt \\
  $^{118}$Kr &  $^{118}$Kr &  $^{264}$Hg &  $^{264}$Hg \\
  $^{120}$Sr &  $^{120}$Sr &  $^{266}$Pb &  $^{266}$Pb \\
  $^{122}$Zr &  $^{122}$Zr &  $^{268}$Po &  $^{268}$Po \\
  $^{130}$Mo &  $^{130}$Mo &  $^{270}$Rn &  $^{270}$Rn \\
  $^{136}$Ru &  $^{136}$Ru &  $^{272}$Ra &  $^{272}$Ra \\
  $^{140}$Pd &  $^{140}$Pd &  $^{274}$Th &  $^{274}$Th \\
  $^{152}$Cd &  $^{152}$Cd &  $^{278}$U  &  $^{278}$U  \\
	     &  	   &  $^{284}$Pu &  $^{282}$Pu \\
\end{tabular}
\label{table02}
\end{ruledtabular}
\end{center} 
\end{table}

Figures \ref{fig07} and \ref{fig08} suggest that the impact of particle-number projection may be instrumental in the formation of sizeable halos, since the RVAP mechanism always guarantees a solution with non-zero pairing correlations. Since the chemical potential is irrelevant in a projected theory we should therefore, in principle, compute this quantity using {\it projected} energies and compare it from the results obtained using un-projected quantities. As emphasized earlier, the application of particle number projection in odd nuclei is not possible at the moment, hence the one neutron RVAP drip line is not accessible. 

We therefore carried out systematic RVAP calculations of the two neutron separation energies,  $S_{2n}$, near the drip line using the D1S interaction. The particular choice of the interaction is secondary in this study, since the focus is on the particular role of particle number projection. The procedure was as follows: for a given drip line element $(Z,N)$ from Table \ref{table01} the isotopes with $N-4$, $N-2$, $N$ and $N+2$ neutrons were considered. For each isotope, the RVAP procedure was carried with $\delta=1.0, 1.05, \dots, 1.50$. The minimum of the RVAP curve was retained as the physical solution for every isotope. The two-neutron separation energy was calculated from the total RVAP-projected energies: $S_{2n} = B^{PNP}(N,Z) - B^{PNP}(N-2,Z)$. The criterion $S_{2n} < 0$ was used to define the position of the new drip line. Table \ref{table02} shows the two-neutron drip line nuclei with and without the particle-number projection.  These two drip lines differ by the isotopes of 6 elements: $^{42}$Mg$_{30}$, $^{62}$Ca$_{42}$, $^{74}$Cr$_{50}$, $^{172}$Sn$_{122}$, $^{262}$Pt$_{184}$ and $^{282}$Pu$_{188}$. As we can read in the neutron number of these nuclei the differences arise always close to the shell closures, where the pairing correlations are either very weak or vanishing.  

\begin{figure}[h]
\includegraphics[height=7.0cm,width=9.0cm]{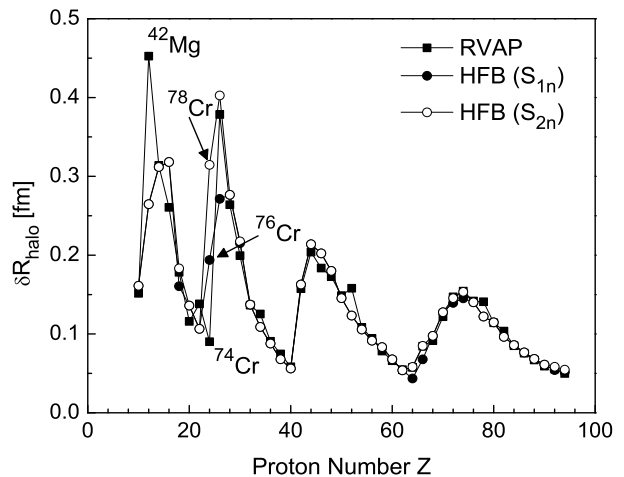}
\caption{Measure of the halo: $\delta R_{halo}(n) = R_{geom}(n) - R_{Helm}(n)$ for RVAP-projected (plain squares), $S_{2n}$-unprojected (open circles) and $S_{n}$-unprojected (plain circles) drip lines. All results are based on spherical Gogny HFB calculations in the WS basis with the D1S interaction. 
	 }
\label{fig09}	
\end{figure}

For all the elements located at the RVAP drip line, the Helm radius was computed, for the protons and the neutrons, based on the projected density $\rho^{PNP}(r)$. The quantity $\delta R_{halo}^{PNP}(n) = R_{geom}^{PNP}(n) - R_{Helm}^{PNP}(n)$ obtained from these calculations is reported in Fig. \ref{fig09}, together with the original $\delta R_{halo}(n)$ of the two un-projected drip lines ($S_{2n}$ and $S_{n}$ drip line).

The impact of particle-number projection is only significant in those nuclei that are unbound at the HFB level but bound in the RVAP-HFB. As can be seen from Table \ref{table02}, there are many nuclei that are particle-unstable ($-S_{1n} \approx \lambda_{n} > 0$) but two-particle-stable ($S_{2n} < 0$). In such cases, the halo is of course larger, sometimes significantly larger like e.g. Cr or Fe, than the corresponding particle-stable isotope. Moreover the value of the halo calculated at the RVAP minimum closely follows the one calculated at the $S_{2n}$ drip line. The case of Cr is singular, in that the RVAP mechanism changes the two neutron drip line by 4 units, thereby considerably lowering the halo. 

However, beyond $Z\approx 30$, the differences between all approaches become relatively negligible. This goes along a very clear and definite trend towards smaller halos as the mass of the nucleus increases. Combining this observation with the fact that our mean-field approach, which includes the continuum and uses the best possible treatment of pairing correlations, fails to produce halos in light nuclei, it is tempting to conclude that halos are a trademark of few-body correlations only. As was recognized early, pairing correlations are a pre-requisite to the formation of halos in a mean-field approach indeed. However, our work seems to further indicate that additional correlations beyond the symmetry conserving mean-field approximation are also mandatory.

Figure \ref{fig08} may suggest that for RVAP solutions the projected density profile is markedly different from the unprojected density. As mentioned already, only the projected density bears a physical meaning by construction. The underlying HFB solution is only used to generate a set of highly pair-correlated projected wave functions. Nevertheless one may compare the behavior of $\delta R_{halo}$ when using either the projected density $\rho^{PNP}(r)$ in the RVAP minimum or the underlying un-projected HFB density $\rho^{HFB}(r)$ in this same minimum to visualize the impact of projection itself. The difference in $\delta R_{halo}$ is practically negligible (less than 0.01 fm) and certainly not on the same scale as differences coming from the interaction. 

We have also applied the RVAP formalism to the calculation of neutron skins and we do not find any remarkable difference as compared with the HFB ones. This confirms the observation that neutron skins are, from a theoretical point of view, mostly sensitive to the details of the interaction (iso-scalar vs. iso-vector content), and from an experimental point of view, to the neutron excess but are not directly affected by the vicinity of the continuum.

In conclusion, we applied our method to include continuum effects in spherical self-consistent HFB calculations with finite-range forces of the Gogny type to the case of nuclear halos and skins. Our calculations show that both the D1 and D1S parametrizations of the Gogny force lead to relatively small mean-field halos, that are of comparable size to most of the results obtained in Skyrme-HFB or relativistic Hartree-Bogoliubov theories. In particular, we do not find the giant halos in Neon and Zirconium isotopes that were reported in several publications. As a rule of thumb we observe that the size of the halo tends to decrease as the mass of the nucleus increases and only light nuclei feature decent-sized halos. By contrast, neutron skins are found to be very clearly related to the ratio (N-Z)/A.

We also show that the impact of particle number projection, before variation, is relatively important since it can change the position of the drip line. 
However, we find that particle-number projected continuum-coupled HFB theory, employing the most realistic form of the pairing interaction, cannot reproduce the large halos observed experimentally in very light nuclei such as $^{11}$Li. This suggests a series of necessary conditions for a successful description of nuclear halos in the framework of mean-field theory: (i) the continuum must be properly included in the formalism (ii) the shell structure must be realistic enough (iii) pairing correlations must be present (iv) symmetry-breaking mean-field calculations, including all relevant deformation degrees of freedom, are probably mandatory (v) all such broken symmetries (in particular particle number) should then be restored (vi) probably configuration mixing such as e.g. GCM should also be included.

{\bf Acknowledgment - } One of us (N.S.) acknowledges financial support of the spanish Ministerio de Educacion y Ciencia (Ref. SB2004-0024). This work has been supported in part by the Spanish Ministerio de Educaci\'on y Ciencia under contract FPA2007-66069, by the Spanish Consolider-Ingenio 2010 Programme CPAN (CSD2007-00042) as well as by the U.S. Department of Energy under Contract Nos. DE-FC02-07ER41457 (University of Washington), DEFG02- 96ER40963 (University of Tennessee), and DE-AC05- 00OR22725 with UT-Battelle, LLC (Oak Ridge National Laboratory). 

%
%

\end{document}